\documentclass{pasa}%

\usepackage{aas_macros}
\usepackage{hyperref} 
\usepackage{xcolor}
\hypersetup{colorlinks,citecolor=blue,linkcolor=blue,urlcolor=blue}
\usepackage{graphicx}
\usepackage{multirow}
\usepackage{amsmath}
\usepackage{mathtools}
\usepackage{subcaption,cleveref}
\usepackage{cprotect}
\usepackage[normalem]{ulem}
\newcommand{\tabft}[1]{(#1)}
\newcounter{ft}
\newcommand{\ft}[1]{\noindent%
  ~\refstepcounter{ft}\tabft{\alph{ft}\label{#1}}}

\def\hour{\ensuremath{^\mathrm{h}}}
\def\minute{\ensuremath{^\mathrm{m}}}
\def\second{\ensuremath{^\mathrm{s}}}
\def\degr{\ensuremath{^\circ}}
\def\arcmin{\ensuremath{^\prime}}
\def\arcsec{\ensuremath{^{\prime\prime}}}
\newcommand\fs{\mbox{\ensuremath{.\!\!^{\mathrm s}}}}%
\newcommand\farcs{\mbox{\ensuremath{.\!\!^{\prime\prime}}}}

\def\spindex{\ensuremath{-1.83\pm0.29}}
\def\mwa{{MWA-2}}

\title[Diffuse, non-thermal radio emission within Abell~1127]{Murchison Widefield Array detection of {steep-spectrum}, diffuse, non-thermal radio emission within Abell~1127}

\author[S.~W. Duchesne et al.]{S.~W. Duchesne$^1$\thanks{email: \url{stefan.duchesne.astro@gmail.com}}, M. Johnston-Hollitt$^{1}$, Z. Zhu$^{2}$, R.~B. Wayth$^{1}$, J.~L.~B. Line$^{1,3}$
\affil{$^1$International Centre for Radio Astronomy Research (ICRAR), Curtin University, Bentley, WA 6102, Australia}%
\affil{$^2$School of Physics and Astronomy, Shanghai Jiao Tong University, 800 Dongchuan Road, Minhang, Shanghai 200240, People's Republic of China}%
\affil{$^3$ARC Centre of Excellence for All Sky Astrophysics in 3 Dimensions (ASTRO 3D)}
}%

\jid{PASA}
\doi{10.1017/pas.\the\year.xxx}
\jyear{\the\year}

\hypersetup{draft}

\begin{document}

\begin{frontmatter}
\maketitle

\begin{abstract}

Diffuse, non-thermal emission in galaxy clusters is increasingly being detected in low-frequency radio surveys and images. We present a new diffuse, steep-spectrum, non-thermal radio source within the cluster Abell~1127 found in survey data from the Murchison Widefield Array (MWA). We perform follow-up observations with the `extended' configuration MWA Phase II with improved resolution to better resolve the source and measure its low-frequency spectral properties. {We use archival Very Large Array S-band data to remove the discrete source contribution from the MWA data,} and from a power law model fit we find a spectral index of {\spindex} broadly consistent {with relic-type sources.} {The source is revealed by the Giant Metrewave Radio Telescope (GMRT) at 150~MHz to have an elongated morphology, with a projected linear size of $850$~kpc as measured in the MWA data.} Using \emph{Chandra} observations we derive morphological estimators and confirm quantitatively that the cluster is in a {disturbed} dynamical state, consistent with the majority of phoenices and relics being hosted by merging clusters. We discuss the implications of relying on morphology and low-resolution imaging alone for the classification of such sources and highlight the usefulness of the MHz to GHz radio spectrum in classifying these types of emission. Finally, we discuss the {benefits and limitations} of using the MWA Phase II in conjunction with other instruments for detailed studies of diffuse, steep-spectrum, non-thermal radio emission within galaxy clusters.

\end{abstract}

\end{frontmatter}

\section{INTRODUCTION }
\label{sec:intro}
Clusters of galaxies are large virialized structures that reside at the intersection of cosmic filaments. The formation of galaxy clusters is thought to be hierarchical through mergers and accretion \citep{pee80}, and cluster mergers represent some of the largest, most energetic collisions in the known Universe. Clusters have been found to host a number of radio emitting sources with steep synchrotron emission spectra implying aged electron populations \citep[see e.g.][]{Pacholczyk1970,Tribble1993,Ensslin01,kbc+04}. Both centrally-located \emph{radio halo} sources and peripherally-located \emph{radio relic} sources are thought to be associated with inter-cluster mergers or otherwise similarly energetic and turbulent events \citep[see][for a review]{vanWeeren2019}. \par
Halos, relics, and other types of steep-spectrum cluster sources have been well-studied over the last two decades. \citet{Ensslin01} propose for radio relics that adiabatic compression of remnant radio galaxy lobes by shocks in the intra-cluster medium (ICM) is responsible for radio relic sources such as that in Abell~85 \citep[][]{Slee1984,Slee01} or Abell~4038 \citep{Slee1998}, however, for megaparsec-scale radio relics a different physical process may be the cause. Diffusive shock acceleration \citep[DSA;][]{Fermi49,Jones1991} is consistent with the spectral features of megaparsec-scale radio relics \citep[see e.g.][]{JohnstonHollitt2003,vanWeeren2010,vanWeeren2016} and shocks have been observed at the locations of some relics. These two types of relics are distinguished as \emph{phoenices} and \emph{radio shocks} \citep[see][but also \citealt{kbc+04}]{vanWeeren2019}. Radio halo type sources can also be broken into two classes: \emph{giant radio halos} and \emph{mini-halos}; the distinction here is less about physical process as both are thought to be caused by turbulence in the ICM through a second order Fermi process \citep{Fermi54,Brunetti2001,Gitti2015}. \par
While the physical (re-)acceleration mechanisms are understood, the origin of the seed electrons for the emission is still not entirely clear. The synchrotron-emitting electrons have been proposed as the thermal pool of electrons in ICM (i.e. the hot plasma, though this has issues related to Coulomb losses; \citealt{Petrosian2001}), left-over from old active galactic nuclei (AGN; see e.g. \citealp{Shimwell2015,vanWeeren2017,deGasperin2017}), or electrons created in  proton-proton collisions \citep{Dennison1980}. The seed for phoenices and mini-halos are almost certainly AGN, despite any difference in final (re-)acceleration process, the larger-scale halos and relics may require a similar origin. \par
As halos, relics, and similar types of emission are usually detected with low surface brightness and steep, mostly power law spectra, radio interferometers such as the  Murchison Widefield Array \citep[MWA;][]{tgb+13} in remote Western Australia are well-suited at detecting and characterising such emission \citep[e.g.][]{hjh+14,gdj+17,djo+17,zjdl18}. In 2017, the GaLactic and Extragalactic All-sky MWA survey \citep{wlb+15,Hurley-Walker2017} was released, unveiling myriad new candidate halos and relics (Johnston-Hollitt et al., in preparation), as well as giving unprecedented spectral coverage to previously detected halos and relics in the Southern Sky. \par
One such galaxy cluster, Abell~1127, was found to host diffuse radio emission within the 72--231~MHz GLEAM survey images, and was found as part of a targeted search for halos and relics within clusters from the Abell catalogues \citep[ACO;][]{aco89}, the Meta-Catalogue of X-ray detected Clusters of galaxies \citep[MCXC;][]{pap+11}, and the \emph{Planck} Sunyaev--Zel'dovich cluster catalogue \citep[PSZ1;][]{planck15}. This particular example is complete with archival X-ray data from the \emph{Chandra} observatory and given its location above the equator has good coverage from northern optical surveys. 
\par 
In this paper we will investigate the diffuse radio source within Abell~1127 (located at $10\hour54\minute14\fs4$, $+14\degr38\arcmin34\farcs8$) and the cluster itself, from radio to X-ray wavelengths to determine its nature. We assume a Lambda Cold Dark Matter cosmology, with $H_0 = 70$~km\,s$^{-1}$\,Mpc$^{-1}$, $\Omega_\mathrm{m} = 0.3$, and $\Omega_\Lambda = 1-\Omega_\mathrm{m}$.

\subsection{Abell 1127 and associated clusters}
\label{sec:abell1127}
Abell~1127 \citep[$10\hour54\minute09\second$, $+14\degr40\arcmin00\arcsec$;][]{aco89} is detected as a \emph{Planck}-SZ source (10$^{\mathrm{h}}$54$^{\mathrm{m}}$18\fs1 $+14$\degr39\arcmin21\arcsec, with positional uncertainty of $\sim2.4$~arcmin; \citealt{psz2}), designated PSZ2~G231.56$+$60.03. A spectroscopic redshift is reported by \citet{whl12} of $z_\mathrm{spec}=0.2994$. However, the cluster is also used as part of the SOAR \footnote{Southern Astrophysical Reasearch telescope} Gravitational Arc Survey \citep[SOGRAS;][]{sogras} wherein a photometric redshift of $z_\mathrm{phot} = 0.328$ is determined.  Despite the difference in redshift we consider this a single cluster system. The system has not previously been described as merging, and no diffuse, non-thermal radio emission has been found to be associated with the cluster. 

\section{DATA}
\label{sec:data}
A plethora of all-sky and large-area astronomical surveys are available, many of which we make use of here. As well as survey data, we have dedicated observations with the MWA and archival data from the \emph{Chandra} X-ray observatory and the {Karl G.~Jansky Very Large Array (VLA)}. We will introduce and describe the various data products in this section. 
\subsection{Surveys}
\subsubsection{Radio surveys}\label{sec:surveys:radio}
{Three radio surveys are used explicitly} in this work: The NRAO \footnote{National Radio Astronomy Observatory} VLA Sky Survey \citep[NVSS;][]{Condon1998}, which covers the entire sky north of $\delta_{\mathrm{J2000}} \geq -40^\circ$ at 1.4~GHz; the Faint Images of the Radio Sky at Twenty-centimeters \citep[FIRST;][]{Becker95,White1997,Helfand2015}, mostly located in the Northern Sky and also at 1.4~GHz, the TIFR GMRT \footnote{Tata Institute for Fundamental Research Giant Metrewave Radio Telescope} Sky Survey Alternative Data Release \citep[TGSS-ADR1;][]{ijmf16}, covering the sky north of $\delta_\mathrm{J2000} \gtrsim -53^\circ$ centred at 150~MHz. {We also utilise low-resolution images from both the NVSS and TGSS ADR1 by convolving full-resolution images to a final resolution of $100~\text{arcsec} \times 100~\text{arcsec}$ to match closely to MWA data.}

\subsubsection{Optical surveys}\label{sec:surveys:optical}
The region surrounding the Abell~1127 system is covered by the Sloan Digital Sky Survey Data III, Data Release 12 \citep[SDSS III, DR12;][]{sdss1,ewa+11,aaa+12}, with 158 sources with redshifts available within 7~Mpc of the diffuse radio source of interest. 

\subsection{VLA S-band data}\label{sec:data:vla}

{In June 2017 Abell~1127 was observed with the VLA in S-band (2--4~GHz) in the C configuration for $\sim24$~min (Proposal ID 17A-308, PI T. Cantwell). S-band is split into 16 subbands of 128~MHz, and each subband is calibrated independently. Calibration and radio frequency interference (RFI) flagging is performed following standard procedures for VLA data using the Common Astronomy Software Applications (\texttt{CASA}) package \citep{casa}. 3C~286 is used for flux scale calibration using the \citet{PerleyButler2017} scale \citep[itself based on the flux density scale of][]{Baars1977}, and 3C~241 is used for phase calibration, bracketing the source observations. Eight subbands are not used due to RFI contamination or calibration problems. The remaining eight subbands are imaged using the widefield imager \texttt{WSClean}  \citep{wsclean1,wsclean2} \footnote{\url{https://sourceforge.net/p/wsclean/wiki/Home/}}, jointly deconvolving the 8 subbands and creating a model in each band which is used for self-calibration with \texttt{CASA}. We perform three rounds of phase-only self-calibration followed by a round of phase and amplitude self-calibration. Data are imaged with a natural weighting, providing the best sensitivity to point sources in this configuration, creating a fullband image centered at 3.063~GHz. No extended emission is seen at the location of the diffuse emission seen in the GLEAM data. We subtract the naturally-weighted model from the visibilities and re-image with a 25~arcsec Gaussian taper to try to highlight extended structure, {finally convolving the resulting image to a final resolution of $100~\text{arcsec} \times 100~\text{arcsec}$}. The naturally-weighted image prior to source subtraction is shown in Fig. \ref{fig:vla:sband}.}

\subsection{MWA data}\label{sec:data:mwa}

\setcounter{ft}{0}
\begin{table*}
    \centering
    \cprotect\caption{\mwa~observational details for Abell~1127.}
    \begin{tabular}{c c c c c c c l}
    \hline
         Date & $N_{88~\mathrm{MHz}}$ \tabft{\ref{tab:field1:obs:nobs}} & $N_{118~\mathrm{MHz}}$ & $N_{154~\mathrm{MHz}}$ & $N_{185~\mathrm{MHz}}$ & $N_{216~\mathrm{MHz}}$ & Project \tabft{\ref{tab:field1:obs:project}} & Comments \tabft{\ref{tab:field1:obs:comment}} \\\hline
         (UTC) & & & & & \\\hline
         2018-01-17 & 13/13 & 14/14 & 14/14 & 12/14 & 14/14 & G0045 & Good. \\
         2018-01-18 & 0/14 & 11/14 & 9/14 & 8/14 & 13/14 & G0045 & Poor ionosphere. \\
         2018-01-26 & 0/6 & 6/6 & 5/5 & 2/5 & 4/5 & G0008 & Variable ionosphere. \\
         2018-02-01 & 0/6 & 0/6 & 2/5 & 2/5 & 4/5 & G0008 & Poor ionosphere. \\
         2018-03-03 & 1/6 & 6/6 & 5/5 & 4/4 & 5/5 & G0008 & Poor ionosphere. \\
         2018-05-07 & 5/5 & 5/5 & 4/4 & 4/4 & 4/4 & G0008 & Good. \\
         2018-05-26 & 5/5 & 5/5 & 4/4 & 4/4 & 2/4 & G0008 & Good. \\
         \hline
         Totals \tabft{\ref{tab:field1:obs:totals}} & 24/55 & 47/56 & 43/51 & 36/50 & 46/51 & ... & ... \\\hline
    \end{tabular} \\
    \footnotesize{\emph{Notes.} \ft{tab:field1:obs:nobs} Ratio of snapshots used to snapshots observed for a given night and frequency, with each snapshot 112--120s of data. \ft{tab:field1:obs:project} G0008 is GLEAM-eXtended, and G0045 is the dedicated cluster project. \ft{tab:field1:obs:totals} Total snapshots used after discarding snapshots with issues including poor ionospheric conditions, too few tiles, or particularly bad RFI. \ft{tab:field1:obs:comment} User determined qualitative assessment of the datasets.}
    \label{tab:field5:obs}
\end{table*}
\setcounter{ft}{0}

\begin{figure}
    \centering
    \includegraphics[width=1\linewidth]{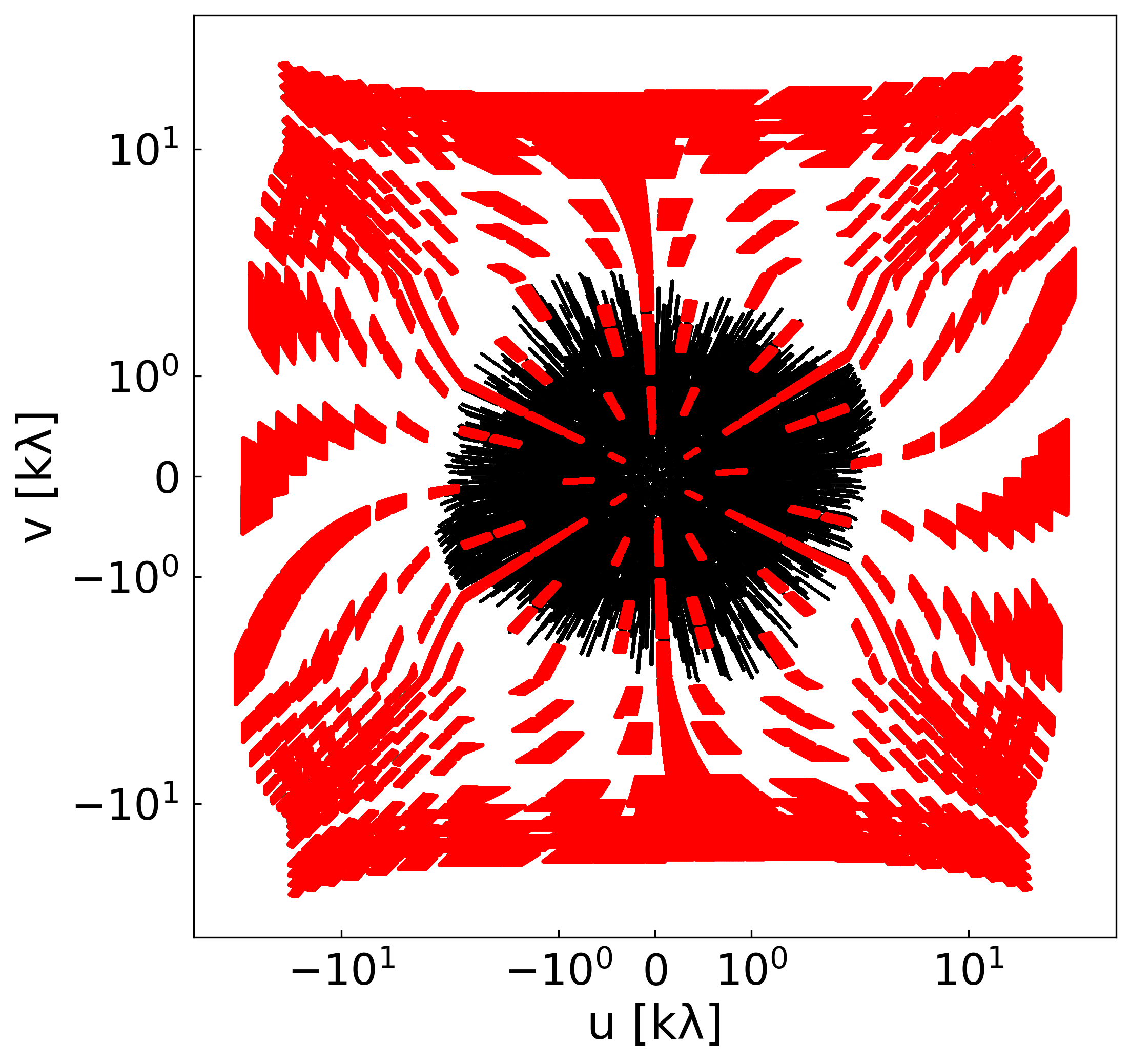}
    \caption{{The typical $u$--$v$ coverage of a single 2-min snapshot at 154~MHz (the central band) with 30~MHz bandwidth (black) with the VLA S-band $u$--$v$ coverage overlaid (red). Note the symmetric logarithmic scale used for both axes to highlight the overlap.}}
    \label{fig:uvcoverage}
\end{figure}

Abell~1127 was observed as part of a MWA project G0045 \footnote{Details of MWA projects can be found at \url{http://www.mwatelescope.org/data/observing}.} during the 2018A observing semester. During this semester, the MWA was operating in the `extended' configuration, part of the Phase II upgrade  \citep[][hereafter \mwa]{wtt+18} that occurred in 2017 adding tiles (antennas) out to $\sim5$~km improving the resolution of the array at a small cost to surface brightness sensitivity. Fig. \ref{fig:uvcoverage} shows the monochromatic $u$--$v$ coverage for a single 2-minute snapshot at 154~MHz illustrating the completeness of the $u$--$v$ coverage with the \mwa. In addition to the G0045 observations, we make use of snapshots taken as part of the \mwa~GLEAM survey (GLEAM-eXtended) where they overlap with the position of A1127 to increase sensitivity. Pertinent observation information is provided in Table \ref{tab:field5:obs}. A set of between 50 and 56 2-minute snapshots for each of the five observing frequencies (at 88, 118, 154, 185, and 216~MHz) were taken across the two projects and across a number of observing nights, though in practice because of poor ionospheric conditions some snapshots were rendered unusable \footnote{Development for direction-dependent calibration of MWA data is underway to help alleviate ionospheric problems, though software and tools to reliably do this are not currently available.}. Over half of the snapshots taken in the lowest frequency band were rendered unusable with current processing techniques. \par
Data processing is largely done using the purpose-written Phase II Pipeline (\verb|piip| \footnote{\url{https://gitlab.com/Sunmish/piip}}) and \verb|skymodel| \footnote{\url{https://github.com/Sunmish/skymodel}} code once data have been pre-processed. The following sections outline this process and the various software involved. 

\subsubsection{Pre-processing}\label{sec:mwa:preprocessing}

Data are recorded by the telescope in 2-minute snapshots as the primary beam varies with time and this allows primary beam correction with current tools. The downside of this observing strategy is that each 2-minute snapshot must be calibrated and imaged individually. Data processing is performed at the Pawsey Supercomputing Centre \footnote{\url{https://pawsey.org.au}} in Perth, Western Australia, which conveniently also hosts the raw visibilities sent directly via fiber from the Murchison Radio-astronomy Observatory (MRO). For each snapshot, the following process is used to generate a calibrated, primary beam corrected image: data are staged using the MWA All Sky Virtual Observatory \footnote{ASVO: \url{https://asvo.mwatelescope.org/}}. The MWA ASVO system converts raw telescope products to the more standard ``MeasurementSet'' format using the \verb|cotter| software developed by A.~O. Offringa, and performs preliminary radio-frequency inference (RFI) flagging using the \verb|AOFlagger| software \citep[][]{ovr12} \footnote{\url{https://sourceforge.net/p/aoflagger/wiki/Home/}}. Additional bad tiles and channels are manually flagged prior to calibration, then again prior to imaging if any bad data present itself. \par

\subsubsection{Initial calibration} \label{sec:mwa:calibration}
As the field of view of the MWA at all frequencies is $>20$ degrees, every snapshot contains a large number of sources which often allows infield calibration, which is made even easier when bright extra-galactic sources (e.g. Virgo A) lie within the main lobe of the primary beam. Generally MWA calibration requires peeling of bright ($> 25$~Jy) sources outside of the image field of view, and in the case of this field the Galactic Plane becomes a source of sidelobe noise at the high end of the MWA band. To counter this, after initial calibration the side lobes of the 185- and 216-MHz bands are imaged  on multiple angular scales and their CLEAN component models (Gaussian and point sources) are subtracted from the visibilities. Note that all CLEAN components within the sidelobe are subtracted, including non-Galactic discrete sources. Despite Virgo A being within and near the field of interest, it either lies within the primary beam mainlobe at low frequencies (88- and 118-MHz) or outside enough to be nulled at higher frequencies (154-, 185-, and 216-MHz). \par
In these observations, we use 100--200 sources for infield calibration, depending on frequency (with more sources used at low frequency due to the larger field view). Calibration is performed with an implementation of the full-Jones \verb|Mitchcal| algorithm, developed for MWA calibration specifically as described by \citet{Offringa2016}, which produces static phase offsets via least-squares fitting for each MWA tile. The sky model used for calibration is generated using a cross-match between GLEAM, NVSS, and TGSS (using the Positional Update and Matching Algorithm, \verb|PUMA|; \citealt{Line2017}) with flux densities and the required frequency estimated by either fitting a generic curved or normal power law to the catalogue flux density measurements, or using an average spectral index of $\langle\alpha\rangle = -0.77$ and assuming a normal power law model. Given the density of measurements provided by the GLEAM Extra-Galactic Catalogue \citep[GLEAM EGC;][]{Hurley-Walker2017}, flux densities are heavily weighted by these measurements {which are based on the \citet{Baars1977} flux density scale, but positions are more heavily weighted by the NVSS positions where available.} \par
\subsubsection{Self-calibration and imaging}
After initial calibration, a single round of phase and amplitude self-calibration is performed by first doing a shallow CLEAN using \texttt{WSClean} which makes use of a \emph{w}-stacking technique to ensure the large field-of-view is properly imaged. This round of CLEAN stops at a threshold of 5 times the local noise. Imaging during self-calibration is done in an ``8 channels out'' mode, where CLEANing is performed on 8 subbands (each with $\Delta\nu = 3.84$~MHz) which allows for a self-calibration model that has frequency-dependence at that sampling. The model generated by \verb|WSClean| is then used by the previously described calibration software to once again calibrate the data. This usually significantly improves the residual calibration artefacts from the first iteration.
Deeper CLEANing is then performed again using \verb|WSClean| with a `Briggs' robust $+1.0$ weighting, this time down to an initial threshold of $3\sigma_{\mathrm{rms}}$ and a final threshold of $1\sigma_\mathrm{rms}$. Here we use a ``4 channels out'' mode (CLEANing subbands of $\Delta\nu = 7.68$~MHz), which addresses the issue of a changing point-spread function (PSF) as a function of frequency, thus reducing amplitude errors which become particularly more problematic around bright sources. Primary beam corrections are applied to each snapshot, using the Full Embedded Element primary beam model \citep{Sokolowski2017}, generating astronomical Stokes \emph{I} primary beam-corrected images. Note that a set of images are also produced at a `Briggs' robust 0.0 weighting but are not used. 

\subsubsection{Astrometric and flux scale corrections}\label{sec:corrections}
For each 2-min snapshot Stokes I image, a pixel-based position correction is done to account for first-order ionospheric effects. This is achieved using \verb|fits_warp.py| \citep{Hurley-Walker2018}, which compares an initial image catalogue generated by the \verb|aegean| \footnote{\url{https://github.com/PaulHancock/Aegean}} source-finding software \citep{Hancock2012,Hancock2018} with a reference catalogue, in this case generated by combining the NVSS, TGSS, and GLEAM catalogues. A pixel-based shift is performed based on the angular separation of sources, and cubic interpolation is used to create an effective screen to shift pixels by. Any snapshots that have significantly higher noise or show more complex ionospheric distortions are at this stage discarded. \par
Finally, each snapshot at each frequency has a slightly different flux scale that must be normalised. The initial calibration is performed with respect to both the GLEAM EGC \citep{Hurley-Walker2017} (along with other radio sky surveys) as well as select multi-component and point-source models from earlier MWA Phase I data of particularly bright sources (e.g. Virgo A). This initial flux scale is more heavily determined by the initial bright source models, which vary from the GLEAM catalogue by up to 50 per cent and so final image-based bootstrapping is required to tie the flux scale more closely to the GLEAM EGC. Additionally, we suspect that there are residual primary beam model errors present in the data, which have a position dependence which is corrected for simultaneously. This process uses in-house code written for this purpose and is described in more detail in Appendix \ref{fluxwarp}. \par

\subsubsection{Stacking the 2-min snapshots}

\begin{figure}
    \centering
    \includegraphics[width=1\linewidth]{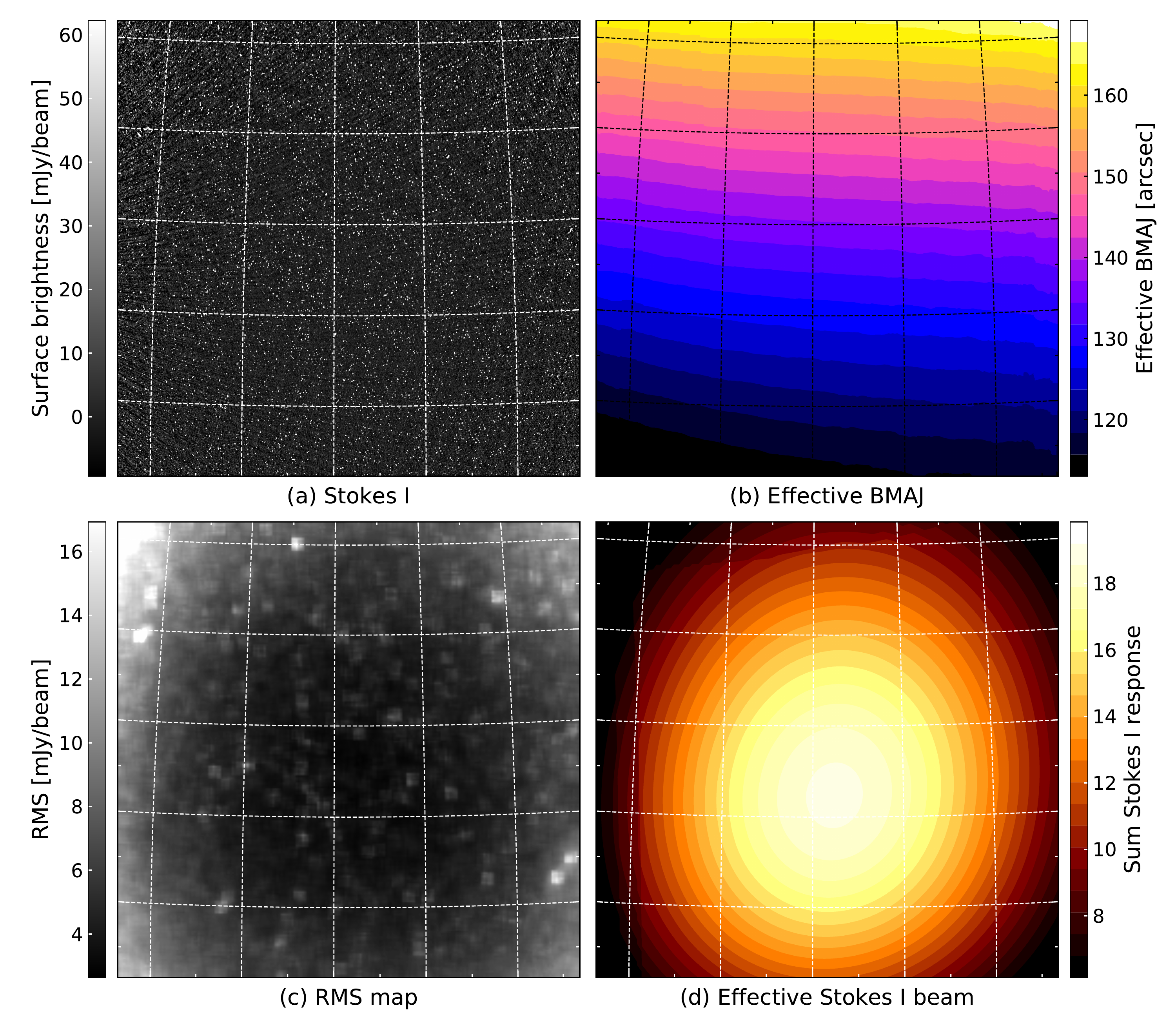}
    \caption{Output from stacking and re-gridding for the 30-MHz band centered on 154~MHz. Note the images cover a $\sim25\degr$ by $\sim 25\degr$ region.}
    \label{fig:stackshots}
\end{figure}

\begin{figure*}
    \centering
    \includegraphics[width=0.9\linewidth]{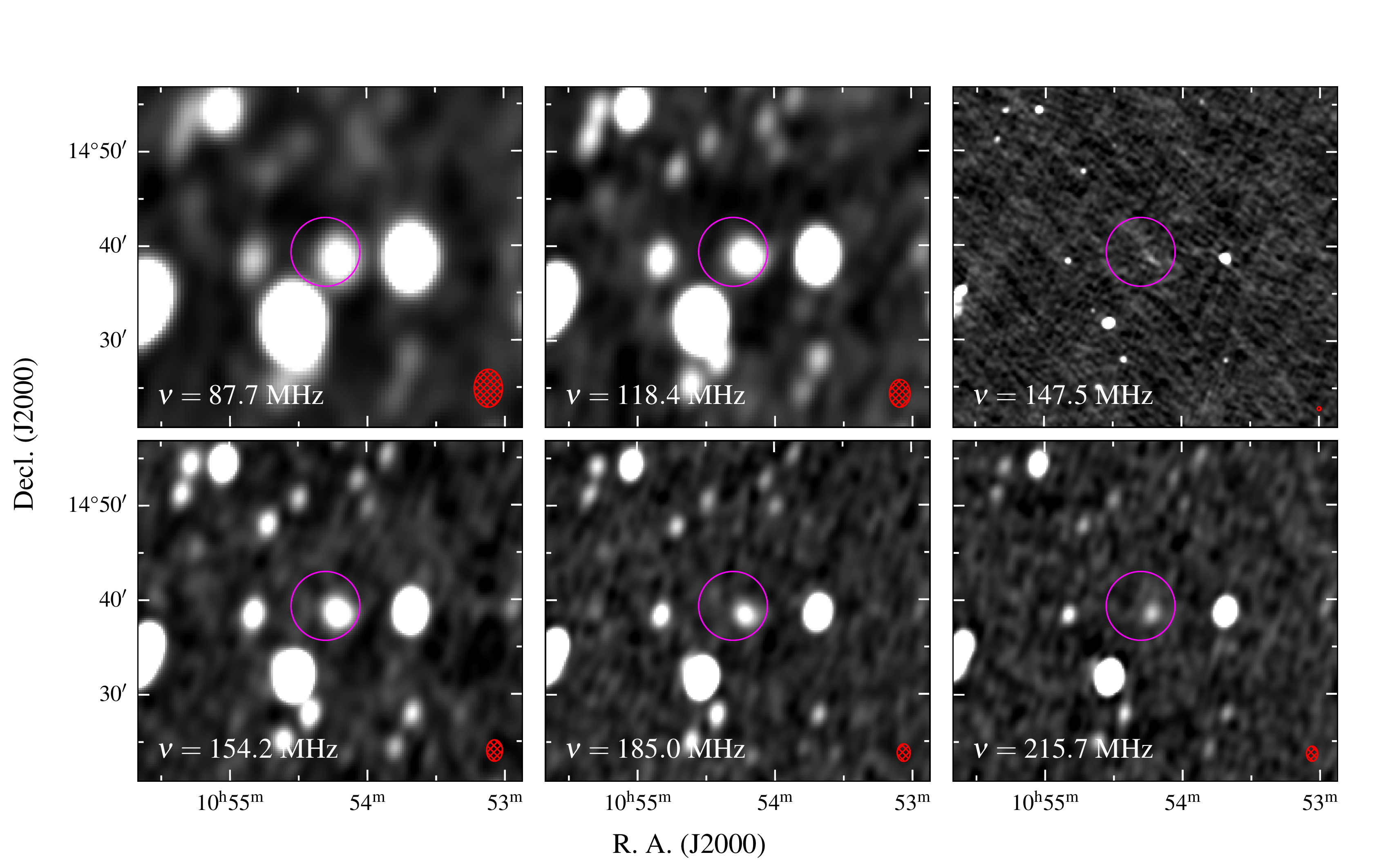}
    \caption{The \mwa~robust $+1.0$ stacked mosaics for the MWA field containing Abell~1127 across the five frequencies, with the TGSS ADR1 image in the top right. The greyscale maps are all linear stretches between $-3\sigma_\mathrm{rms}$ and $15\sigma_\mathrm{rms}$ (see Table \ref{tab:sed} for values for each image). The magenta circle is centered on PSZ2~G231.56$+$60.03 and has a radius of 1~Mpc. The red, hatched ellipses in the lower right corners are the effective PSF size for each map at this position.}
    \label{fig:fieldcomparison}
\end{figure*}

To re-grid images prior to co-addition, we make use of the \verb|regrid| task within the \verb|miriad| software suite \citep{Sault1995}. We generate three separate stacked images: 1) Stokes \emph{I} images weighted by the primary beam response, 2) the effective Stokes \emph{I} response images corresponding to the weighted Stokes \emph{I} image, and 3) the effective PSF map weighted as per the previous two stacked images. In practice, this PSF map incorporates a total flux preserving factor (see Appendix \ref{sec:appendix:psf} for determination of the correction factor) within an effective major axis for the PSF which is all that is required when measuring total or integrated flux densities, hence the position angle is not well defined. Note that peak flux (i.e. surface brightness) is always preserved. Fig. \ref{fig:stackshots} shows the output from creating mosaics for the 154-MHz data, with the Stokes \textit{I} image, effective PSF major axis map, noise map, and summed Stokes \textit{I} primary beam. While our pipeline produces individual snapshots of the $\Delta\nu=30.72$~MHz bands as well as the $\Delta\nu = 7.68$~MHz subbands generated during ``4 channels out'' CLEANing, we only use stacked mosaics of the $\Delta\nu=30.72$~MHz images due to the signal-to-noise ratio constraints of our source of interest in the higher frequency bands (see Section \ref{sec:radio:sed}). The final mosaic properties are presented in Table \ref{tab:phaseIIobs} and the images centered on Abell~1127 are shown in Fig. \ref{fig:fieldcomparison}. Note that some image properties vary over the map (e.g. PSF) so we report the value at the position of Abell~1127. \par

\begin{table}
	\caption{\label{tab:phaseIIobs} Details of the \mwa~observations and resultant images.}
    \resizebox{\linewidth}{!}{\begin{tabular}{c c c c c c c}
		\hline
        Band & $\nu_{\mathrm{c}}$ \tabft{\ref{tab:ref:nu}} & $t_\mathrm{scan}$ \tabft{\ref{tab:ref:t}} & PSF \tabft{\ref{tab:ref:beam}} & $\sum A_I$ \tabft{\ref{tab:ref:sumI}} & $\sigma_\mathrm{rms}$ \tabft{\ref{tab:ref:rms}} & $\Delta_\mathrm{flux}$ \tabft{\ref{tab:ref:fluxunc}} \\
        \hline
        (MHz) & (MHz) & (min) & ($\arcsec \times \arcsec$) & & (mJy\,beam$^{-1}$) & \% \\
        \hline
         72--103 & 88 & 38 & $241.1 \times 179.6$ & 10.1 & 18.4 & 5.6 \\
        103--134 & 118 & 82 & $176.4 \times 131.2$ & 18.1 & 7.1 & 3.5 \\
        139--170 & 154 & 74 & $136.2 \times 101.0$ & 18.0 & 3.7 & 3.7 \\
        170--200 & 185 & 60 & $111.3 \times 82.5$ & 15.1 & 3.4 & 4.1 \\
        200--231 & 216 & 92 & $96.7 \times 71.5$ & 10.6 & 3.9 & 5.0 \\
        \hline
    \end{tabular}} \\
    \footnotesize{\emph{Notes.} \\
    \ft{tab:ref:nu}~Central observing frequency. \\
    \ft{tab:ref:t}~Total scan length of data used in imaging. \\ \ft{tab:ref:beam}~Effective PSF at the source location. \\
    \ft{tab:ref:sumI}~The summed Stokes \emph{I} primary beam response of the stacked image, where $A_I=1$ is the peak response for a 2-min observation at zenith. \\
    \ft{tab:ref:rms} Local rms at the location of Abell~1127. \\
    \ft{tab:ref:fluxunc} \% uncertainty on the flux scale compared to the input calibration model.}
\end{table}

\begin{table}
    \centering
    \caption{Sources detected in the SDSS images at the center of the cluster region.}
    \begin{tabular}{c l c}
         \hline
         ID & Name & $z$ \\\hline
         A1 & GALEXASC~J105418.23$+$143902.3 & - \\
         A2 & SDSS~J105418.12$+$143902.0 & - \\
         B & 2MASX~J10541751$+$1439041 & 0.2994 \\
         C & 2MASX~J10541735$+$1439012 & - \\
         D & SDSS~J105415.58$+$143914.8 & - \\
         E & 2MASS~J10541703$+$1438353 & - \\
         \hline
    \end{tabular}
    \label{tab:SDSSsources}
\end{table}

\subsection{\emph{Chandra} data}
A1127 (as RM~J105417.5$+$143904.2) was observed with the Advanced CCD Imaging Spectrometer (ACIS-I) instrument on the \textit{Chandra} observatory with 10.46~ks of exposure time (Obs. ID 17160, PI: Eduardo Rozo). These archival data were retrieved from the \textit{Chandra} Data Archive (CDA). We analyzed the data obtained from the S0-3 chips of ACIS. We followed the standard \textit{Chandra} data reduction process and used the \verb|CIAO| \footnote{\emph{Chandra} Interactive Analysis of Observations} (version v4.11 with CALDB v4.8.2; \citealt{ciao}) script \verb|chandra_repro| to generate the level-2 event file. We examined the light curves extracted in 0.5--12.0 keV from source-free regions near CCD edges and exclude the time intervals during which the count rates deviate from the mean values by 20 per cent. The CIAO tool \verb|celldetect| is used to identify and exclude the point sources detected on the S0-3 chips and finally \verb|flux_image| is used to generate the exposure map to correct for the vignetting and exposure time fluctuations.

\section{RESULTS}
\subsection{The optical and X-ray core}

\begin{figure*}
    \centering
    \begin{subfigure}{0.5\linewidth}
    \includegraphics[width=1\linewidth]{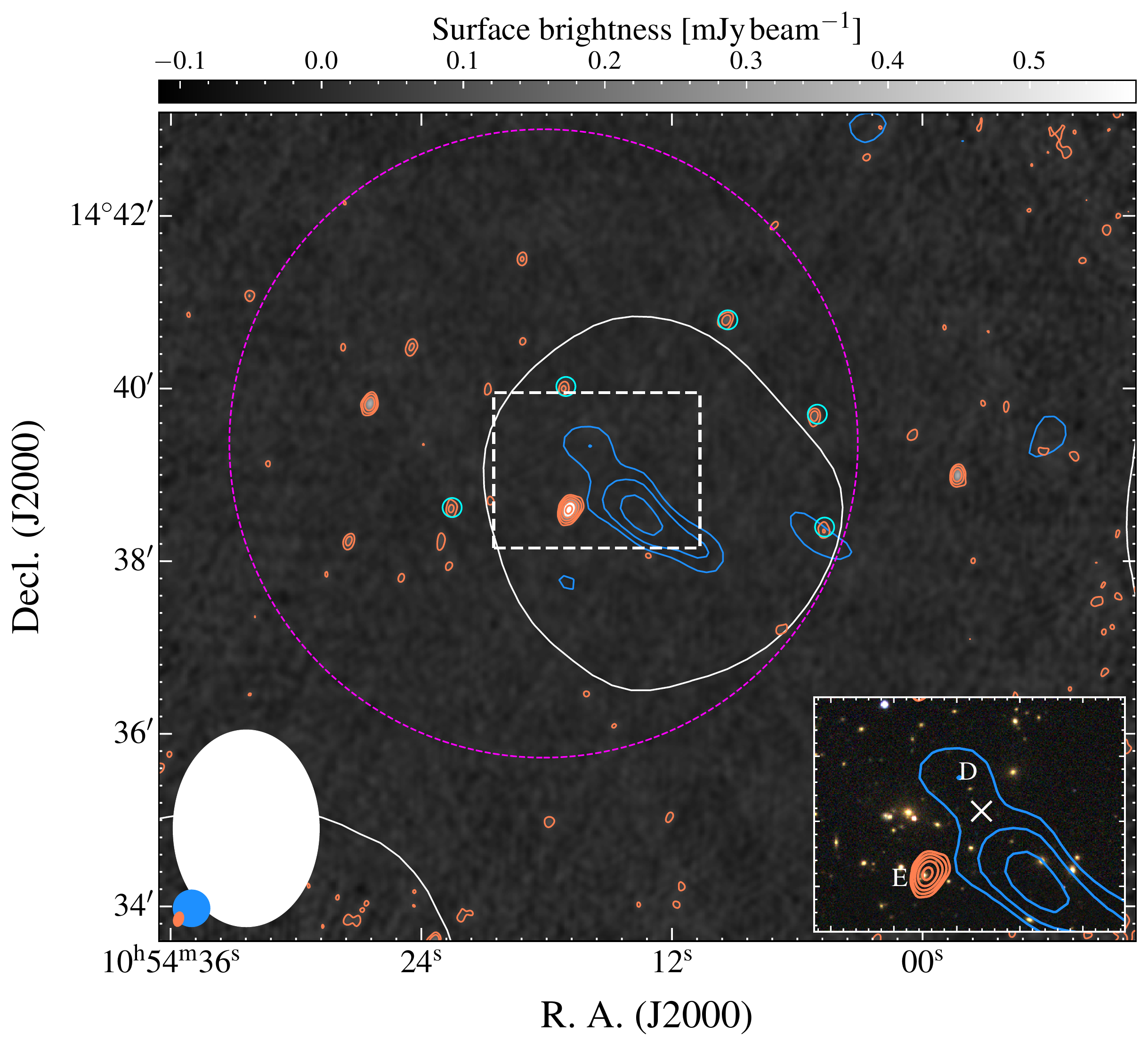}
    \caption{\label{fig:vla:sband}}
    \end{subfigure}%
    \begin{subfigure}{0.5\linewidth}
    \includegraphics[width=1\linewidth]{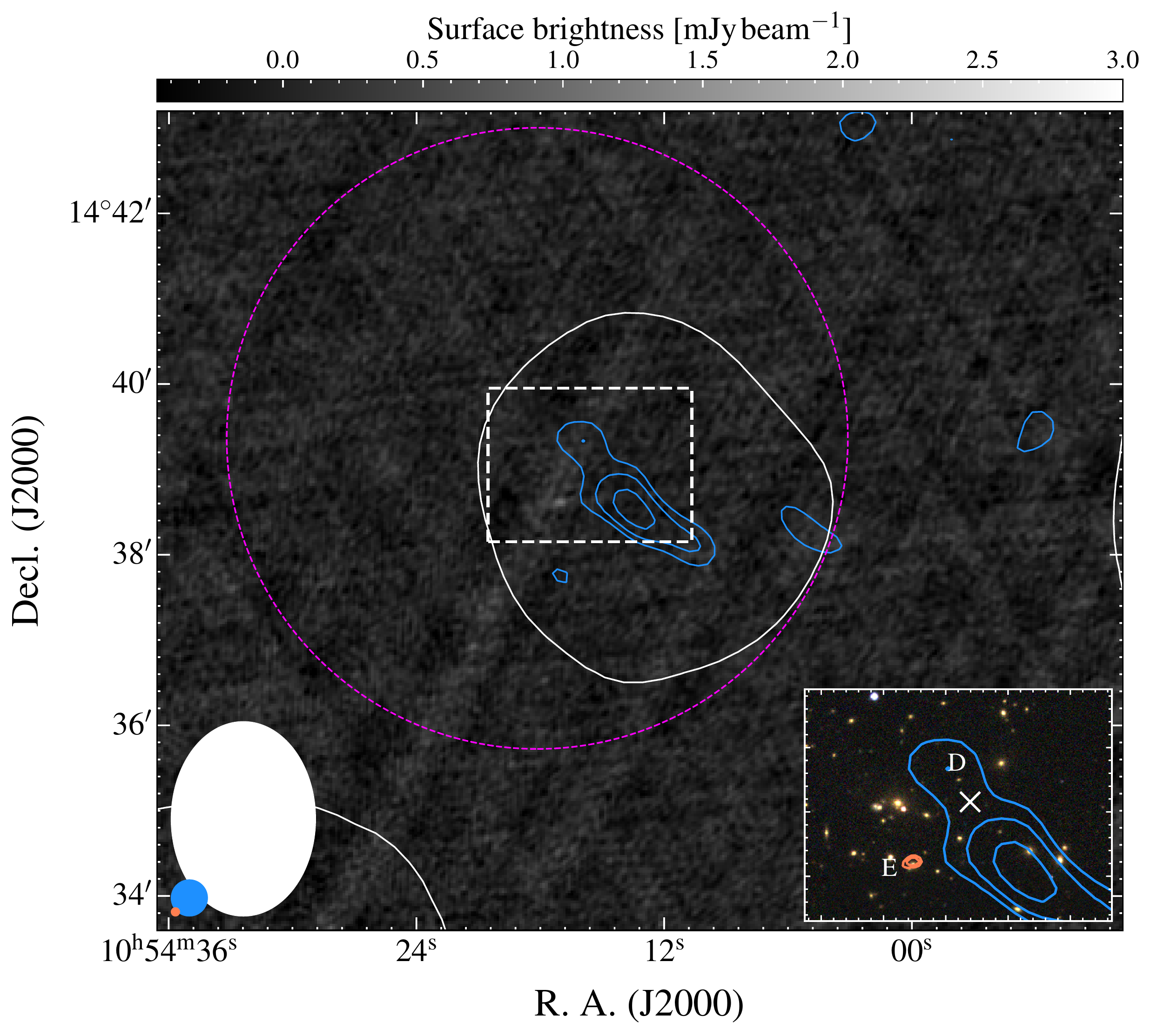}
    \caption{\label{fig:vla:first}}
    \end{subfigure}%
    \caption{\label{fig:vla} 
    {VLA GHz view of the cluster region. \subref{fig:vla:sband} The background is the 3.063~GHz VLA S-band map prior to source-subtraction. Orange contours start at $4\sigma_\mathrm{rms}$ {($\sigma_\mathrm{rms} = 11.5$~$\mu$Jy\,beam$^{-1}$)} increasing with factors of 2. The cyan circles denote discrete sources that affect the MWA measurements as discussed in the main text. \subref{fig:vla:first} FIRST survey image background of the same region. In both panels the magenta circle is as in Fig. \ref{fig:fieldcomparison}. The single white contour is the 154-MHz \mwa~data at $3\sigma_\mathrm{rms}$, and the blue contours are the TGSS data starting at $3\sigma_\mathrm{rms}$ increasing with factors of $\sqrt{2}$. The coloured ellipses in the lower right of each panel are of the respective beam shapes, with smallest, orange beam being the VLA data. The dashed, white box indicates the location of the inset panels. Sources ``E'' and ``D'' are discussed in the text.} 
    }
\end{figure*}

\begin{figure}
    \centering
    \includegraphics[width=1\linewidth]{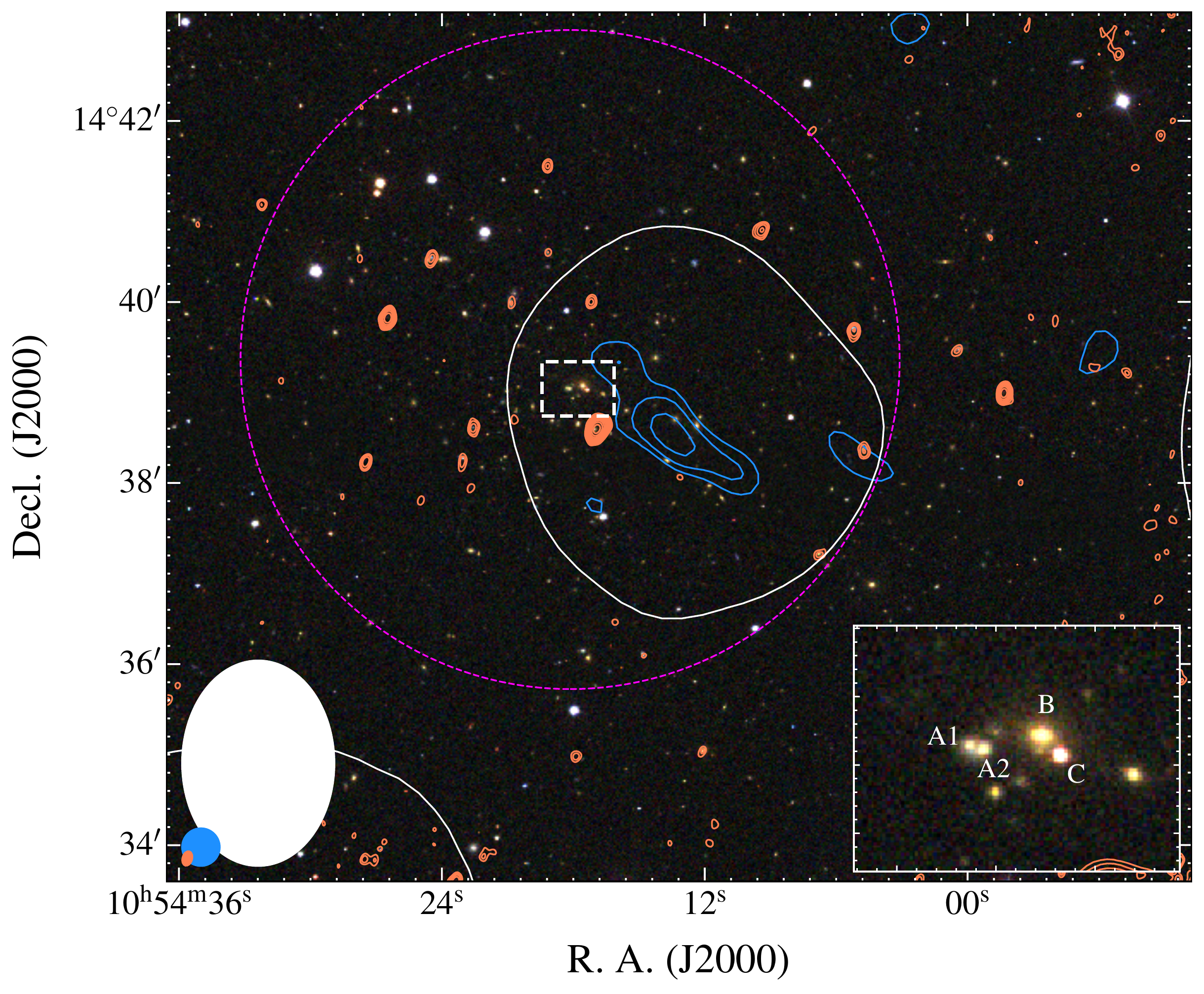}
    \caption{\label{fig:sdss} {The background is a three-colour (red-green-blue) image made from the $i$, $r$, and $g$ bands of the SDSS. The contours and features in the image are as in Fig. \ref{fig:vla:sband}, though the inset location is focused on the cluster centre. The dashed box in the centre of the cluster is enlarged in the inset shown on the bottom right.}}
\end{figure}

{The core of the cluster system can be characterised by both an optical concentration of galaxies as well as an X-ray--emitting plasma. Fig. \ref{fig:sdss} shows the SDSS data of the cluster region, with an inset zoom-in on the BCG and surrounding galaxies (``A1-2'', ``B'', and ``C'').  The BCG, ``B'', is 2MASX~J10541751$+$1439041 (hereafter 2MASX~J1504). 2MASX~J1504 is reported with a spectroscopic redshift of $z=0.299437\pm0.000054$ from DR12 of the SDSS.}

\begin{figure}
    \centering
    \includegraphics[width=1\linewidth]{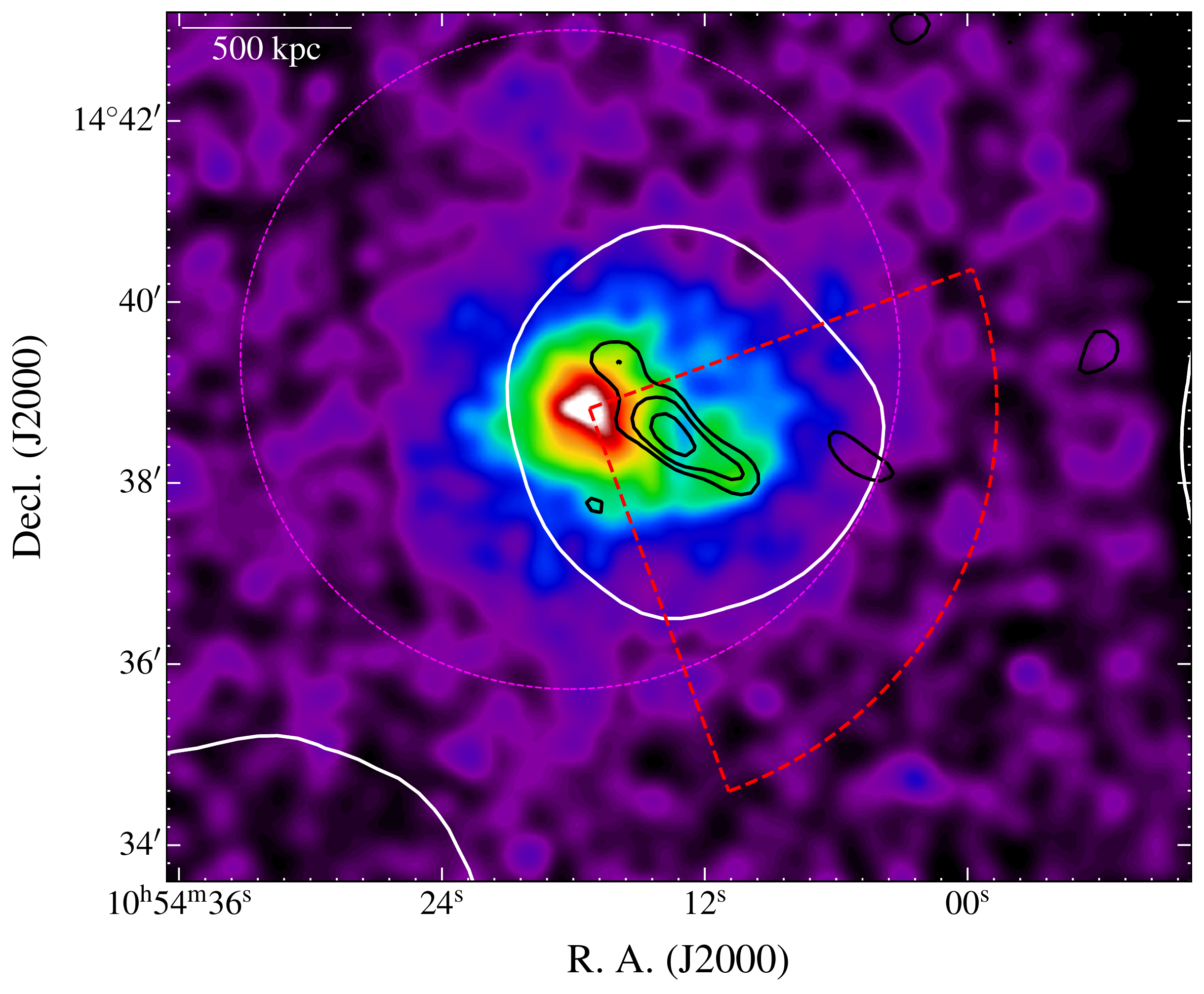}
    \caption{The \emph{Chandra} X-ray map smoothed with a $\sigma=18$~arcsec Gaussian kernel. {The overlaid white contour is from the 154-MHz \mwa~image at $3\sigma_\mathrm{rms}$. The overlaid black contours are the 150-MHz TGSS ADR1 image, with contours also starting at $3\sigma_\mathrm{rms}$. The dashed, magenta circle is the same as in Fig. \ref{fig:fieldcomparison}, and the dashed, red wedge indicates the region used to extract the surface brightness profile through the southwest X-ray clump (blue points in Fig. \ref{fig:xraysb})}.}
    \label{fig:xray}
\end{figure}

 Fig. \ref{fig:xray} shows the reprocessed archival \emph{Chandra} data  with radio contours overlaid. For display purposes we smooth the X-ray image by convolving with a $\sigma=18$~arcsec Gaussian kernel using the \verb|CIAO| task \verb|asmooth|. The image itself provides two interesting things to note: 1) the X-ray distribution is not circular, and 2) the emission is divided into two clumps, with the main, eastern clump containing the peak of the surface brightness and the secondary, western clump being much fainter. The void between the clumps coincides with the peak of the radio emission. 
 
 \begin{figure}
    \centering
    \includegraphics[width=1\linewidth]{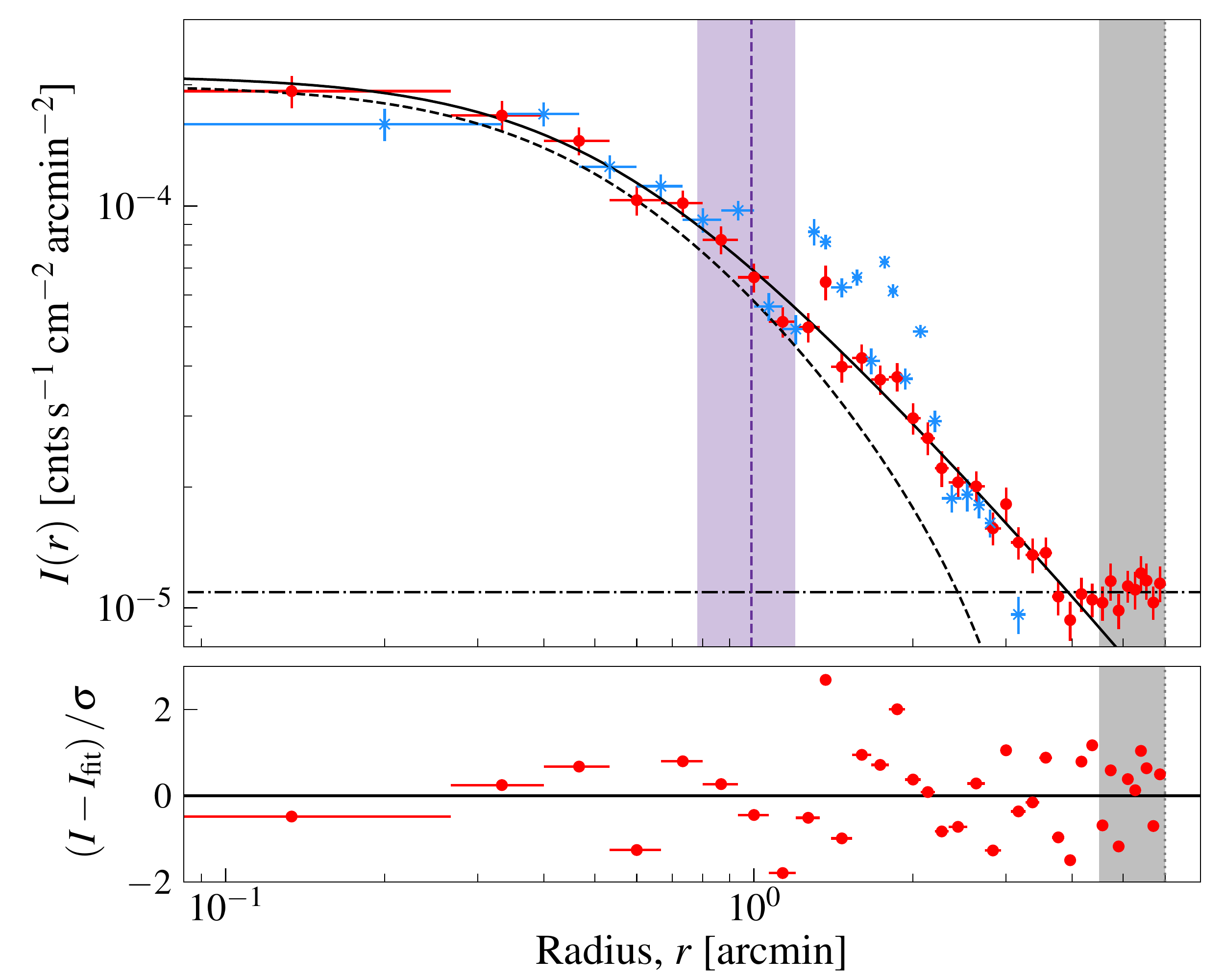}
    \caption{Surface brightness profile of the X-ray emission associated with Abell~1127. For the azimuthally averaged profile (red circles), two models are fit at separate radii: a $\beta$ model for $r \leq 4.5\arcmin$ (dashed line) fit to background subtracted data, and a constant model for $ 4.5\arcmin < r \leq 6\arcmin$ (dot-dash line) to determine the background. The solid line is the combination of the background-subtracted $\beta$ model and the constant background. The grey, shaded region indicates the background fitting region. The horizontal bars indicate the radial bin widths. {The blue points correspond to a radial surface brightness profile across the southwest X-ray clump (red, dashed region in Fig. \ref{fig:xray}), and the vertical, dashed purple line marks the peak emission of the radio source in the TGSS ADR1 image, with the purple shaded region indicating the beam size.}}
    \label{fig:xraysb}
\end{figure}
 
We analyse the un-smoothed \emph{Chandra} image with the X-ray surface brightness analysis software, \texttt{proffit} \footnote{\url{http://www.isdc.unige.ch/~deckert/newsite/Proffit.html}} \citep{Eckert2011}. We use circular annuli to determine an azimuthally averaged surface brightness profile. As the un-smoothed image has no well-defined peak, we use the smoothed image to define the centre of the surface-brightness profile (corresponding to coordinates 10$^{\mathrm{h}}$54$^{\mathrm{m}}$17\fs3, $+14$\degr38\arcmin49\arcsec). The profile is measured out to 6~arcmin (1.6~Mpc at $z_\mathrm{spec}=0.2994$, however this radius is chosen as it is the edge of the image), with counts initially binned in 4~arcsec annuli. These bins are adjusted to ensure a signal-to-noise ratio of at least 10 per bin, resulting in 8--16~arcsec bins with $>100$~counts per bin. We assume a constant background over the image, and define the annuli with radii $4.5\arcmin < r \leq 6\arcmin$ to consist of only background counts from visual inspection and fit these bins with a constant profile. This is $I_\mathrm{b} = (1.095 \pm 0.034) \times 10^{-5}$~counts\,s$^{-1}$\,cm$^{-2}$\,arcmin$^{-2}$ and is subtracted from the surface brightness profile. For the background-subtracted annuli with radii $r \leq 4.5\arcmin$, we find that a standard $\beta$ model \citep{Cavaliere1976} fits the surface brightness profile well (with $\chi_\mathrm{red} = 1.27$). The results of the surface brightness profile fitting and background subtraction are shown in Fig. \ref{fig:xraysb}. {We perform a similar surface brightness analysis across the southwest X-ray clump as indicated by the red, dashed wedge region in Fig. \ref{fig:xray}. This profile is shown in Fig. \ref{fig:xraysb} and the location of the peak radio emission in the TGSS ADR1 image is also plotted for reference. The peak radio emission occurs immediately as the X-ray separates into the southwestern clump.}

 We calculate the centroid shift, $w$ \citep[e.g.][but see also \citealt{Mohr1993}]{Poole2006} with an outer radius set to 1.87~arcmin, corresponding to 500~kpc \citep[see][]{Cassano2010}, resulting in $w_{500}=0.072$. {For further comparison to literature data, we estimate the centroid shift within $R_{500}$ \footnote{{$R_{500}$ corresponds to the radius within which the mean mass density is 500 times the critical density of the Universe.}}. We estimate $R_{500}\sim920$~kpc from a 0.5--2.0~keV X-ray luminosity of $L_\text{X} \sim 3.6\times10^{44}$~erg\,s$^{-1}$ as measured from flux within the cluster region, using $R_{500}$--$L_\text{X}$ relations \citep[][but see also \citealt{Arnaud2005}]{Bohringer2007}; we find $w_{920} = 0.02R_{500}$. } Additionally, we calculate the surface brightness concentration parameter \citep[see e.g.][]{Santos2008} in two ways: within 100~kpc and 500~kpc (0.37 and 1.87~arcmin, respectively, as per \citealt{Cassano2010}) and within 40~kpc and 400~kpc (0.15 and 1.50~arcmin, respectively, as per \citealt{Santos2008}), thus finding $c_{100/500} = 0.105$ and $c_{40/400} = 0.022$. 
 
\subsection{Radio emission}\label{sec:radio}

\subsubsection{Radio morphology and discrete sources}\label{sec:radio:morphology}

The {full-resolution} TGSS ADR1 image at 150~MHz shows an elongated radio structure near the optical centre of the cluster system (see Fig. \ref{fig:sdss}). {Given the concentration of optical sources, it is difficult to confirm if {one or more of the optical galaxies in the region is the host of} the emission, however, no discrete radio sources are seen within the TGSS-detected emission either in the FIRST survey image or the VLA S-band image (Fig.~\ref{fig:vla}). Optical source ``D'' (Fig.~\ref{fig:sdss}) sits within the TGSS emission and at first glance appears to correspond to a peak in the TGSS image with $S_\mathrm{150~MHz} \sim 12$~mJy\,beam$^{-1}$. {However, no discrete source is detected in the VLA S-band image above $3\sigma_\mathrm{rms}$ ($\sigma_\mathrm{rms} = 11.5$~$\mu$Jy\,beam$^{-1}$) at this position. If this component of the full extended emission is a discrete source, it would require $\alpha < -1.9$ to result in a non-detection in the VLA S-band image. Given this would be an unusually steep spectrum core, we assume the peak 150-MHz emission co-spatial with ``D'' is not associated with ``D'' and is part of the extended emission.}}

{The VLA S-band data reveal six discrete radio sources, including Source ``E'', within the measured region in the \mwa~images. Source ``E'' is shown in the insets of Fig. \ref{fig:vla} and other discrete sources are indicated by cyan circles in Fig. \ref{fig:vla:sband}, though the additional discrete sources are not detected in the FIRST survey image. Source ``E'' is not detected in the TGSS image above 9.4~mJy ($3\sigma_\mathrm{rms}$) but based on the spectral index between the FIRST and VLA S-band data ($\alpha_\mathrm{E} = -0.8 \pm 0.2$) , assuming a power law, the source would be just detectable at $\sim 4\sigma_\mathrm{rms}$ significance in the TGSS image. }

{{We measure the full extent of the diffuse source in the MWA-2 154-MHz image: the deconvolved largest angular size is determined to be 3.2~arcmin measured out to $2\sigma_\mathrm{rms}$, corresponding to a projected linear size of $850$~kpc.}}

\subsubsection{Radio spectral energy distribution and power}\label{sec:radio:sed}
We measure the spectral energy distribution of the source between {88--3063}~MHz using the \mwa, TGSS, NVSS, and VLA S-band data. {We measure the integrated flux density of the source in the \mwa~and TGSS data by integrating over {a circular aperture centered on the source}. We mask pixels below 2$\sigma_\mathrm{rms}$, and the measurement uncertainty, $\sigma_S$, is defined via} \begin{equation}
    \sigma_S =  \sqrt{\dfrac{\Omega_\mathrm{pixel} \langle \Omega_\mathrm{beam}\rangle}{N_\mathrm{pixel}}} \displaystyle \sum_{i=0}^{N_\mathrm{pixel}} \dfrac{\sigma_{\mathrm{rms},i}}{\Omega_{\mathrm{beam},i}} \quad [\mathrm{Jy}]\, ,
\end{equation}
where $\Omega_\mathrm{pixel}$ is the constant pixel solid angle, $\Omega_\mathrm{beam}$ is the varying beam solid angle, and $\sigma_\mathrm{rms}$ is the map rms. While the synthesized beam does vary in size across these \mwa~mosaics, in practice the variation across our source of interest is minute. Note that the 216-MHz band of the \mwa~data is considered a lower limit as the shape of the source changes significantly enough that we suspect the entirety of the source is not detected in this band. Inspecting the highest frequency subband ($\nu_\mathrm{c} = 227$~MHz) shows only a hint of signal at the 3$\sigma_\mathrm{rms}$ level. 

\setcounter{ft}{0}
\begin{table}
    \centering
    \caption{Flux density measurements of the diffuse radio source {with required corrections.}}
     \resizebox{\linewidth}{!}{\begin{tabular}{c c c c c c c}
        \hline
         Band & $\nu_\mathrm{c}$ & $S_\nu$ & $S_\mathrm{bias}$ &  $\Delta S_\mathrm{discrete}$ & $S_\mathrm{E}$ & $\sigma_\mathrm{rms}$ \\\hline
         & (MHz) & (mJy) & (mJy) & (mJy) & (mJy) & (mJy\,beam$^{-1}$) \\\hline
        TGSS \tabft{\ref{tab:beam}} &  147.5 & $150 \pm 50$ & - & 6 & $ < 12.2 $ & 25 \\
        NVSS \tabft{\ref{tab:beam}} & 1400 & $< 15$ & - & 0 & $ 1.85 \pm 0.24 $ & 1.2  \\
        S-band \tabft{\ref{tab:beam}} & 3063 & $< 0.9 $ & - & 0 & $0.96 \pm 0.05$  & 0.082  \\
        \hline\multicolumn{7}{c}{\mwa}\\\hline
        \,\,72--103 & 87.7 & $341 \pm 54$ & 48 & 9 & 19 &  18 \\
        103--134 & 118.4 & $188 \pm 29$ & 20 & 7 & 15 & 9 \\
        139--170 & 154.2 & $122 \pm 19$ & 10 & 6 & 12 & 5 \\
        170--200 & 185.0 & $86 \pm 15$ & 12 & 5 & 10 & 4 \\
        200--231 & 215.7 & $>55$ & 14 & 5 & 9 & 4 \\
        \hline
    \end{tabular}}
    {\footnotesize \textit{Note.} \ft{tab:beam} Based on low-resolution image ($100~\text{arcsec} \times 100~\text{arcsec}$).}
    \label{tab:sed}
\end{table}

\begin{figure}
    \includegraphics[width=1\linewidth]{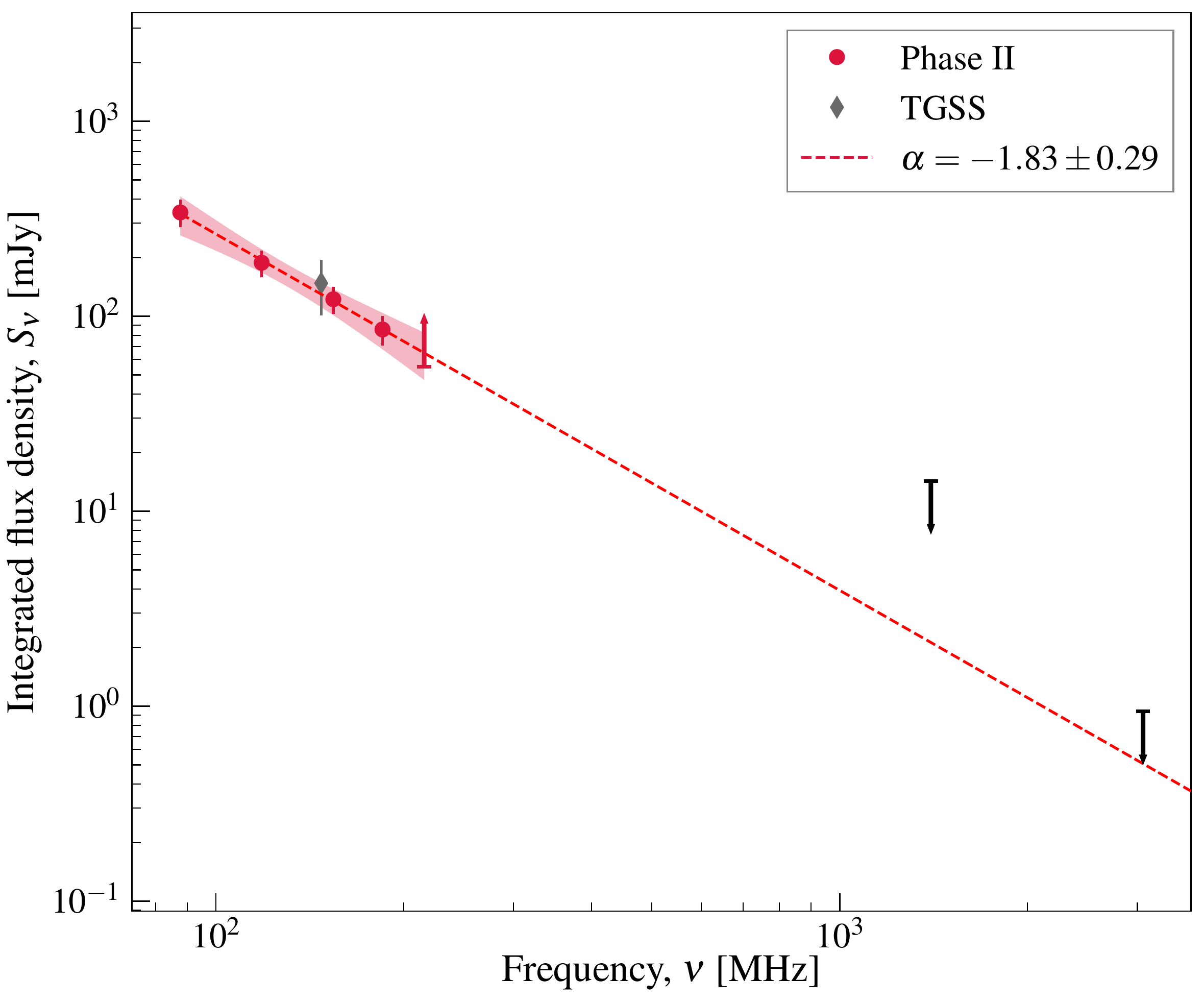}
	\caption{\label{fig:sed} {The spectral energy distribution of the diffuse radio source from {88--3063}~MHz. The black upper limits {are from the low-resolution NVSS and VLA source-subtracted S-band images}. Note the lower limit at 216-MHz. The power law fit to the \mwa~data is shown as the dashed, red line. Limits are not used in fitting. {The TGSS measurement is also shown for completeness, but not used in fitting.} The shaded region corresponds to the 95\% confidence interval.}}
\end{figure}

{For the \mwa~data, we correct a CLEAN bias by adding 14--48 mJy to each measurement (see Appendix \ref{appendix:bias} for details). Additionally, we estimate the possible contribution from the unnamed discrete S-band sources by extrapolating to MWA frequencies assuming $\alpha_\mathrm{discrete} = -0.77$. We find the contribution is small (at the level of the noise) so include this as additional uncertainty in the measurements. The total uncertainty of an integrated flux density measurement is the quadrature sum of the measurement uncertainty, flux scale uncertainty, and discrete source uncertainty. The contribution from source ``E'' is also estimated from $\alpha_\mathrm{E} \sim -0.8$ and is subtracted. {The low-resolution TGSS ADR1 flux density measurement is consistent with the \mwa~measurement at 154~MHz within the estimated uncertainties. Finally, assuming the angular size of the emission in the MWA-2 154-MHz map is the true size, we estimate a 3$\sigma$ upper limit from the low-resolution NVSS and VLA discrete-source--subtracted S-band maps which have constant rms noises of 1.25 and 0.082~mJy\,beam$^{-1}$, respectively. Table \ref{tab:sed} presents the flux densities measurements and the various measurement corrections for each band.}}

We fit a generic power law model to the \mwa~data between {88--185}~MHz using the Levenberg--Marquardt algorithm for non-linear least-squares fitting implemented in \verb|lmfit| \citep{lmfit}. The best-fit power law model yields a spectral index of {$\alpha = \spindex$} for the source. This fit is shown in Fig. \ref{fig:sed}.
Using the \mwa~model fit we estimate the 1.4-GHz flux density of the extended source to be {$S_\mathrm{1.4~GHz} \sim 2$~mJy}, which gives a monochromatic power of {$P_\mathrm{1.4~GHz} \sim 7\times 10^{23}$~W\,Hz$^{-1}$} if indeed the emission is at the redshift of the BCG ($z=0.2994$), and using the spectral index determined above. \par

\section{Discussion}\label{sec:discussion}

\subsection{A dynamic system}\label{sec:discussion:merging}

{The clumping of the X-ray emission and general extension to the west suggests an un-relaxed system}. We can attempt to quantify this by comparing the morphological parameters $c_{100/500}=0.105$ ($c_{40/400} = 0.022$) and $w_{500}=0.072$ ({$w_{920}=0.02R_{500}$}) and  with other clusters. The surface brightness concentration parameter is a good indicator of cool-core systems \citep{Santos2008} and here we note that $c_{40/400}=0.026$ is below values typically seen in cool-core systems (with $c_{40/400} \gtrsim 0.75$). \citet{Poole2006} find that for simulated data the centroid shift, $w$, is a good indicator that a cluster has been disturbed, presumably by merger-related activity. \citet{Pratt2009} define a disturbed system as having {$w > 0.01R_{500}$} from a representative sample of 31 X-ray--emitting, nearby clusters (i.e. the REXCESS \footnote{Representative \textit{XMM-Newton} Cluster Structure Survey} sample), suggesting that Abell~1127 is morphologically disturbed. Additionally, the measured values for $w_{500}$ and $c_{100/500}$ place the cluster in the quadrant of merging clusters in Figure 1(a) from \citet{Cassano2010}, most of which have been found to host giant radio halos. {The \emph{Chandra} data provide good support for the cluster system being in a morphologically disturbed state (corresponding to merger activity)}. {In addition to this, the peak emission of the radio source in the TGSS ADR1 image sits immediately before the transition between the main X-ray--emitting clump and the fainter southwestern clump.}

\subsection{Classification of the diffuse radio source}\label{sec:discussion:classification}
{While there are numerous optical galaxies within the emission region of the extended, diffuse radio emission there is no radio core detected. This, combined with the steep observed radio spectrum precludes the extended source from being a normal, active radio galaxy. The steep radio spectrum of the source suggests an aged population of electrons, fading or perhaps re-accelerated from merger-related activity within the ICM. Merging clusters have been found to host radio halos and relics \citep[see e.g.][]{Cassano2013}. While the source is unlikely to be a giant radio halo, given its offset from the X-ray emission peak, we consider the possibility of a relic-like radio source.}

{Unfortunately the source is not resolved enough with the current data to perform a resolved spectral study to explore possible radio shock origins \citep[e.g.][]{vanWeeren2010,hjh+14,deGasperin2014} including re-acceleration or re-energisation \citep[e.g.][]{Bonafede2014,deGasperin2017}, however, if the source is a relic generated from a shock, it is likely oriented at some angle between the cluster and observer, and shock-driven relic  features such as a spectral gradients may not be present or observable. The integrated spectrum is steeper than most radio relics associated with shocks (with the current sample mean $\alpha = -1.2 \pm 0.2$; \citealt{vda+19} and references therein), but does share observed properties of the relic source in RXC~J1234.2$+$0947 \citep{Kale2015}; a similar steep spectrum relic-like source with no observed connection to a shock. Such steepness is more often seen in ``roundish'' radio relics or phoenices, thought to be energised by adiabatic compression due to small-scale shocks \citep{kbc+04}, possibly from cluster mergers. We do not rule out a merger-related relic classification based on the present data.}

{An alternative explanation is that of remnant (non--re-accelerated) electrons from a long-dead radio galaxy---confirmation of this would require, at the least, access to a higher-frequency detection of the emission to confirm spectral steepening \citep[see e.g.][]{Murgia2011,Duchesne2019}. Potential hosts for such a scenario are sources ``B'' (the BCG) or ``E'', with ``E'' the most likely candidate based on existing detected emission at 1.4~GHz. In either case this requires some separation of the radio lobes from the host and, assuming a maximum projected velocity of $~1000$~km\,s$^{-1}$ (away from the diffuse radio source) requires a travel time of $\gtrsim 200$~Myr.}

{While we may speculate on the nature of the emission, from the data at hand it is impossible to confirm its precise classification.}

\subsection{Towards an SED-based taxonomy: current limitations}\label{sec:discussion:taxonomy}

{SED sampling is sorely missing in many studies of diffuse, non-thermal radio cluster emission. Some examples exist of well-sampled spectra, though it is only the brightest examples of diffuse cluster emission, such as the radio halo in the Coma Cluster \citep[e.g.][]{Schlickeiser1987,Thierbach2003}, or the relic-type source in Abell~85 \citep{Slee01} or Abell~4038 \citep[][]{Slee01,Kale2018}, that have well-studied and sampled spectra which allow the distinction between synchrotron emission models for the sources. The MWA provides good fractional bandwidth at MHz-frequencies, however, cannot be used alone to fully confirm integrated emission models. For completeness, additional data at GHz-frequencies would be required to distinguish between various emission model with breaks or curves in logarithmic space between MHz and GHz frequencies.}

{The most significant limit of MWA data (even in its Phase II `extended' configuration) is the angular resolution. This is limiting for two reasons: 1) the intrusion of discrete sources within the larger-scale cluster emission, and 2) the limited ability to perform resolved spectral studies. The first limitation can be bypassed by incorporating complementary higher-resolution observations or survey data as used in this work. In the near future, the Australian Square Kilometre Array Pathfinder \citep[ASKAP;][]{askap1,askap2} will be providing the Evolutionary Map of the Universe \citep[EMU;][]{emu1} survey, covering the Southern Sky up to $+30^\circ$ declination, complementing the coverage offered by the MWA. At a frequency of $\sim 900$~MHz, resolution of 10~arcsec, and expected noise of 10~$\mu$Jy\,beam$^{-1}$, we will be able to provide better analysis of intruding discrete sources than what is provided here with the current VLA S-band data and FIRST survey images. Additionally, where there is overlap between the MWA and TGSS (and by extension the the newly upgraded GMRT; \citealt{Gupta2017}) we can immediately rule out bright compact sources within the emission or that make up the emission. The second limitation is not bypassable with the current array, but still resolved spectral studies can be performed on the nearest, largest sources \citep[e.g.][]{hjh+14}. For resolved spectral studies, a combination of instruments such as the LOw Frequency ARray \citep[LOFAR;][]{lofar}, (u)GMRT, and VLA have been used to good effect with deep observations \citep[e.g.][]{deGasperin2017,DiGennaro2018}, noting the appropriate caveats in regards to matching $u$--$v$ coverage.}

\section{SUMMARY}
\label{sec:conclusion}

{In this paper we have presented observations of a steep-spectrum, diffuse radio source in the cluster Abell~1127. The data include dedicated \mwa~observations, and archival VLA (S-band) and \emph{Chandra} data, as well as survey data from the TGSS, FIRST, and NVSS, and SDSS. With the available data, we are unable to unambiguously classify the radio source, but we report on the following properties: \begin{enumerate}
    \item steep radio spectrum, $\alpha=\spindex$, up to GHz-frequencies,
    \item projected linear extent {$850$~kpc},
    \item hosting cluster is morphological disturbed,
    \item no obvious radio core.
\end{enumerate}
These features are consistent with radio relics, phoenices, as well as remnant radio galaxies, and places this in a growing category of similar diffuse cluster sources which are not able to be precisely categorised with present data \citep[e.g.][and a number of references within \citealt{vanWeeren2019}]{Shakouri2016,djo+17}.}

{We have described a data-reduction pipeline for \mwa~continuum data based on the pipeline used for the GLEAM survey, with improvements to the calibration and overall flux scale and improvements to how the re-projected PSF is handled during image stacking/mosaicking.}

{Despite additional long baselines provided by the \mwa~``extended'' configuration, MWA data are still limited in angular resolution ($\gtrsim 50$~arcsec). We have showed that despite this limiting angular resolution, with complementary high resolution observations to remove discrete source contribution, we can investigate diffuse cluster sources. In the near future, the sky observable to the MWA will have complementary $\sim 10$~arcsec resolution data with sufficient sensitivity to disentangle underlying point source populations as well as the necessary surface-brightness sensitivity to detect diffuse cluster sources.}

\begin{acknowledgements}
The authors would like the thank Davide Picchieri and Susannah R. Keel for identifying the source of interest. {The authors would also like to the thank the anonymous referee for their useful comments that have improved the quality of this paper.}\par
SWD acknowledges an Australian Government Research Training Program scholarship administered through Curtin University. ZZ is supported by the National Science Foundation of China (grant No.~11433002, 11835009). \par
This scientific work makes use of the Murchison Radio-astronomy Observatory, operated by CSIRO. We acknowledge the Wajarri Yamatji people as the traditional owners of the Observatory site. Support for the operation of the MWA is provided by the Australian Government (NCRIS), under a contract to Curtin University administered by Astronomy Australia Limited. \par
We acknowledge the Pawsey Supercomputing Centre which is supported by the Western Australian and Australian Governments. \par
This research has made use of data obtained from the Chandra Data Archive and software provided by the Chandra X-ray Center (CXC) in the application package \verb|CIAO|. \par
{The National Radio Astronomy Observatory is a facility of the National Science Foundation operated under cooperative agreement by Associated Universities, Inc.}
This research made use of a number of \texttt{python} packages: \texttt{aplpy} \citep{Robitaille2012}, \texttt{astropy} \citep{astropy:2013,astropy:2018}, \texttt{matplotlib} \citep{Hunter2007}, \texttt{numpy} \citep{Numpy2011} and \texttt{scipy} \citep{Jones2001}. \par
This research has made use of the NASA/IPAC Extragalactic Database (NED), which is operated by the Jet Propulsion Laboratory, California Institute of Technology, under contract with the National Aeronautics and Space Administration. \par
Funding for SDSS-III has been provided by the Alfred P. Sloan Foundation, the Participating Institutions, the National Science Foundation, and the U.S. Department of Energy Office of Science. The SDSS-III web site is \url{http://www.sdss3.org/}.
SDSS-III is managed by the Astrophysical Research Consortium for the Participating Institutions of the SDSS-III Collaboration including the University of Arizona, the Brazilian Participation Group, Brookhaven National Laboratory, Carnegie Mellon University, University of Florida, the French Participation Group, the German Participation Group, Harvard University, the Instituto de Astrofisica de Canarias, the Michigan State/Notre Dame/JINA Participation Group, Johns Hopkins University, Lawrence Berkeley National Laboratory, Max Planck Institute for Astrophysics, Max Planck Institute for Extraterrestrial Physics, New Mexico State University, New York University, Ohio State University, Pennsylvania State University, University of Portsmouth, Princeton University, the Spanish Participation Group, University of Tokyo, University of Utah, Vanderbilt University, University of Virginia, University of Washington, and Yale University.
\end{acknowledgements}

\bibliographystyle{pasa-mnras}
\bibliography{reference}

\begin{appendix}

\section{The effective point spread function}\label{sec:appendix:psf}
The effective PSF is not well defined in images after regridding and reprojecting with current software (e.g. \texttt{regrid} from \texttt{miriad}; \citealt{Sault1995}, or \texttt{SWarp} \citealt{swarp}). This problem is exacerbated by a shift of reference coordinates over tens of degrees, and by the large field of view of the MWA. To ensure the reprojected PSF is defined correctly for integrated flux density measurements, we define an effective PSF correction factor, $f_\mathrm{regrid}$, dependent on final projection, to determine the effective PSF area. This factor is
\begin{equation}\label{eq:fregrid}
f_\mathrm{regrid} = \begin{dcases}
\sqrt{\dfrac{1 - l^2 -m^2}{1 - l^{\prime 2} - m^{\prime 2}}} & \text{if }\texttt{SIN} \, , \\
\sqrt{1 - l^2 -m^2} & \text{if }\texttt{ZEA} \, ,
\end{dcases}
\end{equation}
where $l,m$ are the direction cosines with respect to the original image reference coordinates and $l^\prime,m^\prime$ with respect to the new image reference coordinates. Note that this assumes the original images are in a \texttt{SIN} projection as output by \texttt{wsclean}. For the work here we have final reprojected images in the \texttt{SIN} projection, however, it is common with MWA data to also produce mosaics with the \texttt{ZEA} projection \citep[see e.g. GLEAM;][]{Hurley-Walker2017}. Naturally the \texttt{ZEA} correction only takes the \texttt{SIN} component from the original image due to the equal area definition of the \texttt{ZEA} projection. \par
An additional concern regarding the PSF and measuring integrated flux densities was found in the \texttt{aegean} source-finding software. Prior to commit \texttt{6cd5bac} \footnote{\url{https://github.com/PaulHancock/Aegean/commit/6cd5bac42405a654c26f43d6971b893444fdd1c7}} calculation of the PSF size was done assuming a distortion due to projection and, optionally, with an additional latitude-dependent correction if not in the \texttt{SIN} projection. These factors produced an approximate correction that became worse radially from the image reference coordinates. Removing these factors and applying $f_\mathrm{regrid}$ to the effective PSF area (when data have been reprojected) produces the expected results. \par

\begin{figure*}
    \centering
    \begin{subfigure}{0.45\linewidth}
    \includegraphics[width=1\linewidth]{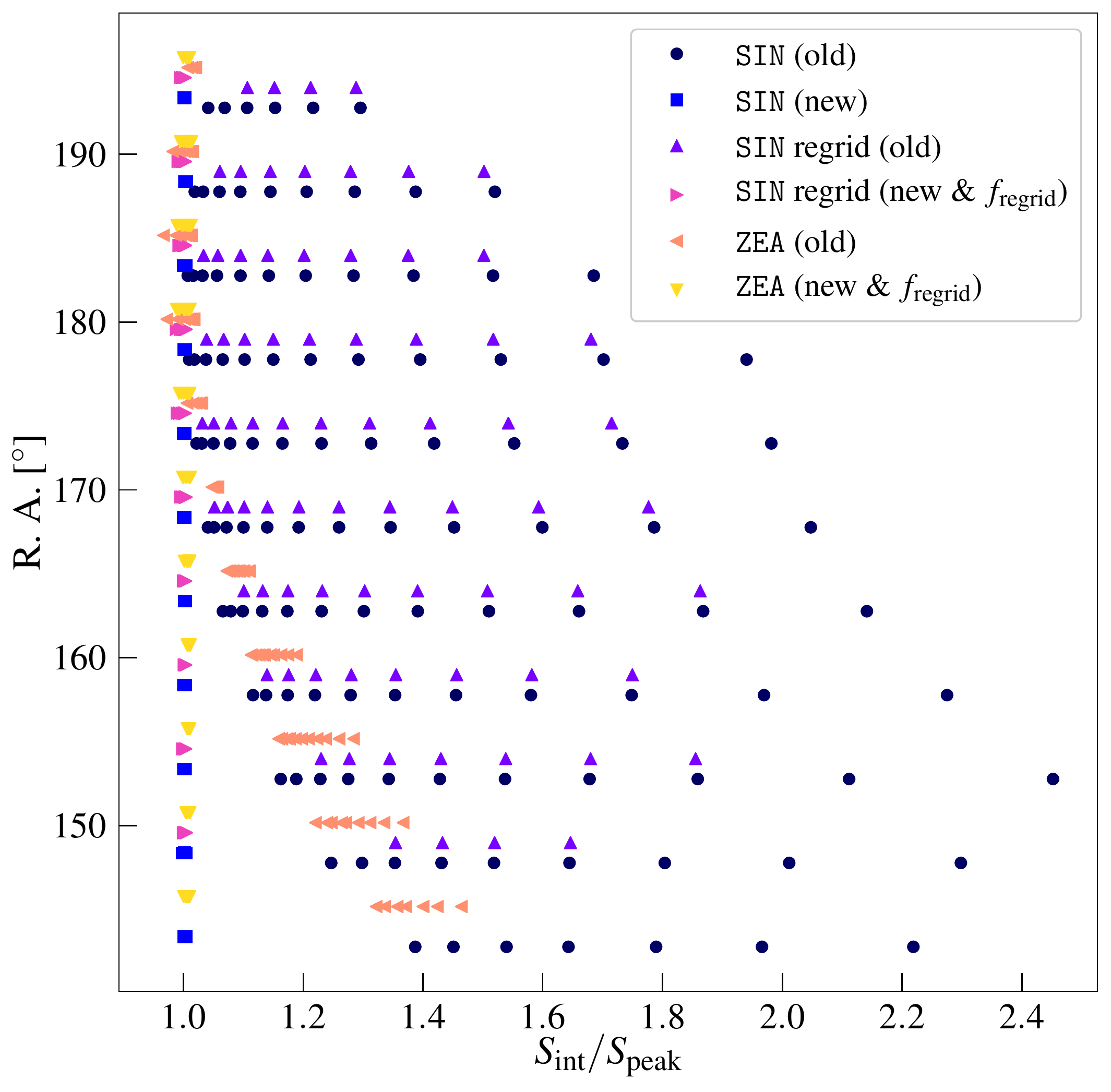}
    \caption{\label{fig:appendix:psf:highdec:ra} Low-elevation pointing, RA dependence.}
    \end{subfigure}\hfill%
    \begin{subfigure}{0.45\linewidth}
    \includegraphics[width=1\linewidth]{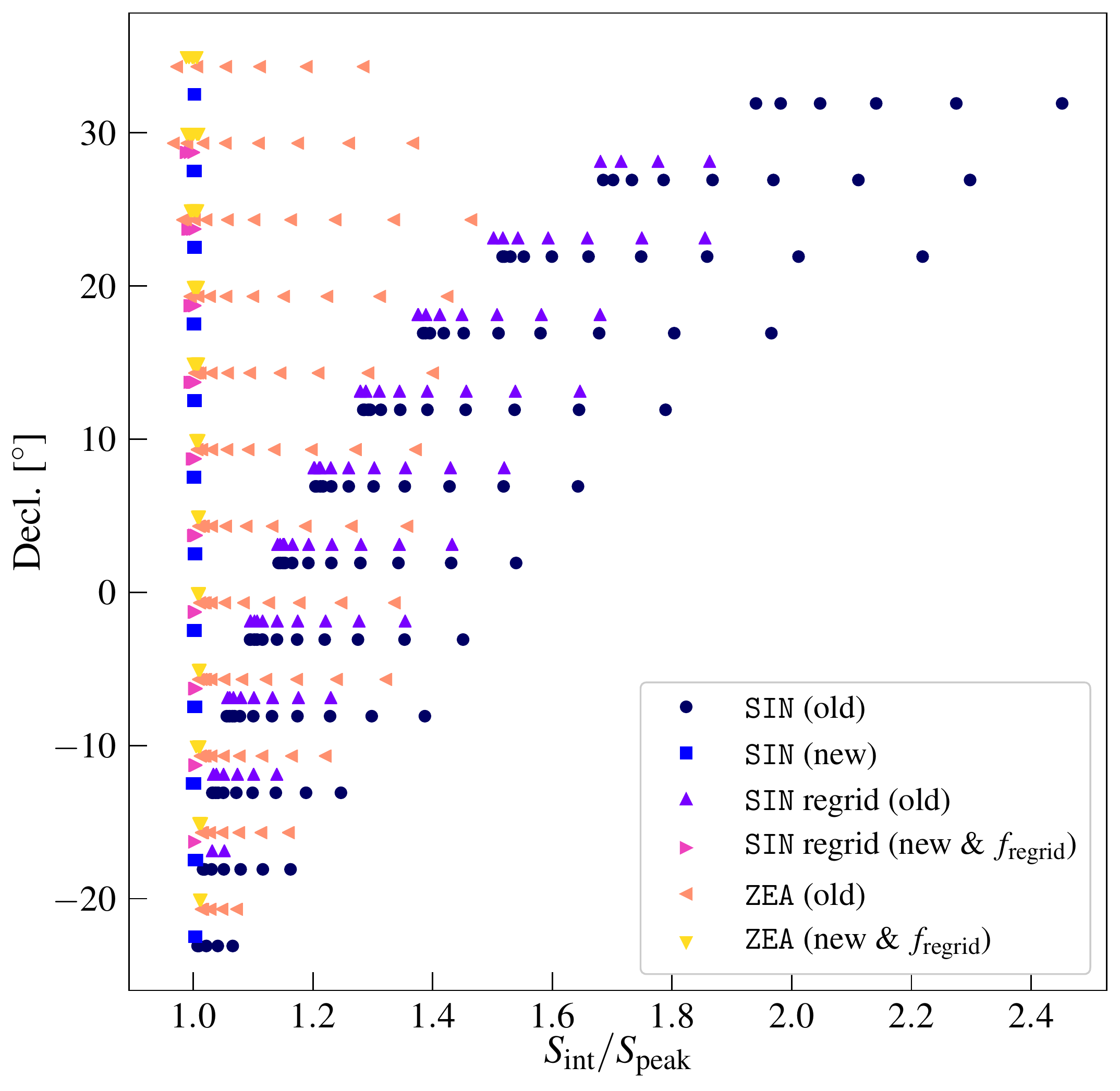}
    \caption{\label{fig:appendix:psf:highdec:dec} Low-elevation pointing, declination dependence.}
    \end{subfigure}\\~\\%
    \begin{subfigure}{0.45\linewidth}
    \includegraphics[width=1\linewidth]{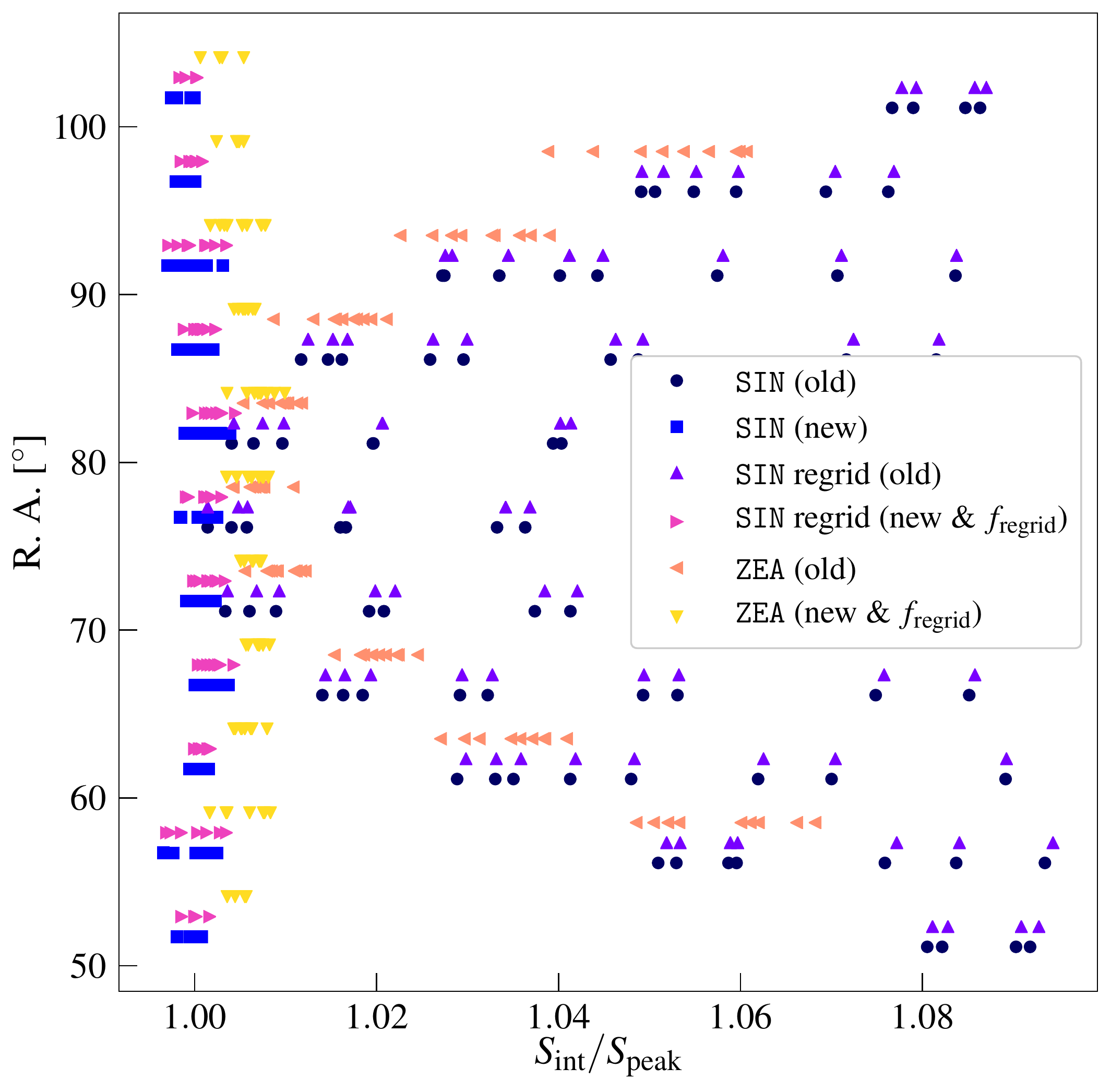}
    \caption{\label{fig:appendix:psf:zenith:ra} Zenith pointing, RA dependence.}
    \end{subfigure}\hfill%
    \begin{subfigure}{0.45\linewidth}
    \includegraphics[width=1\linewidth]{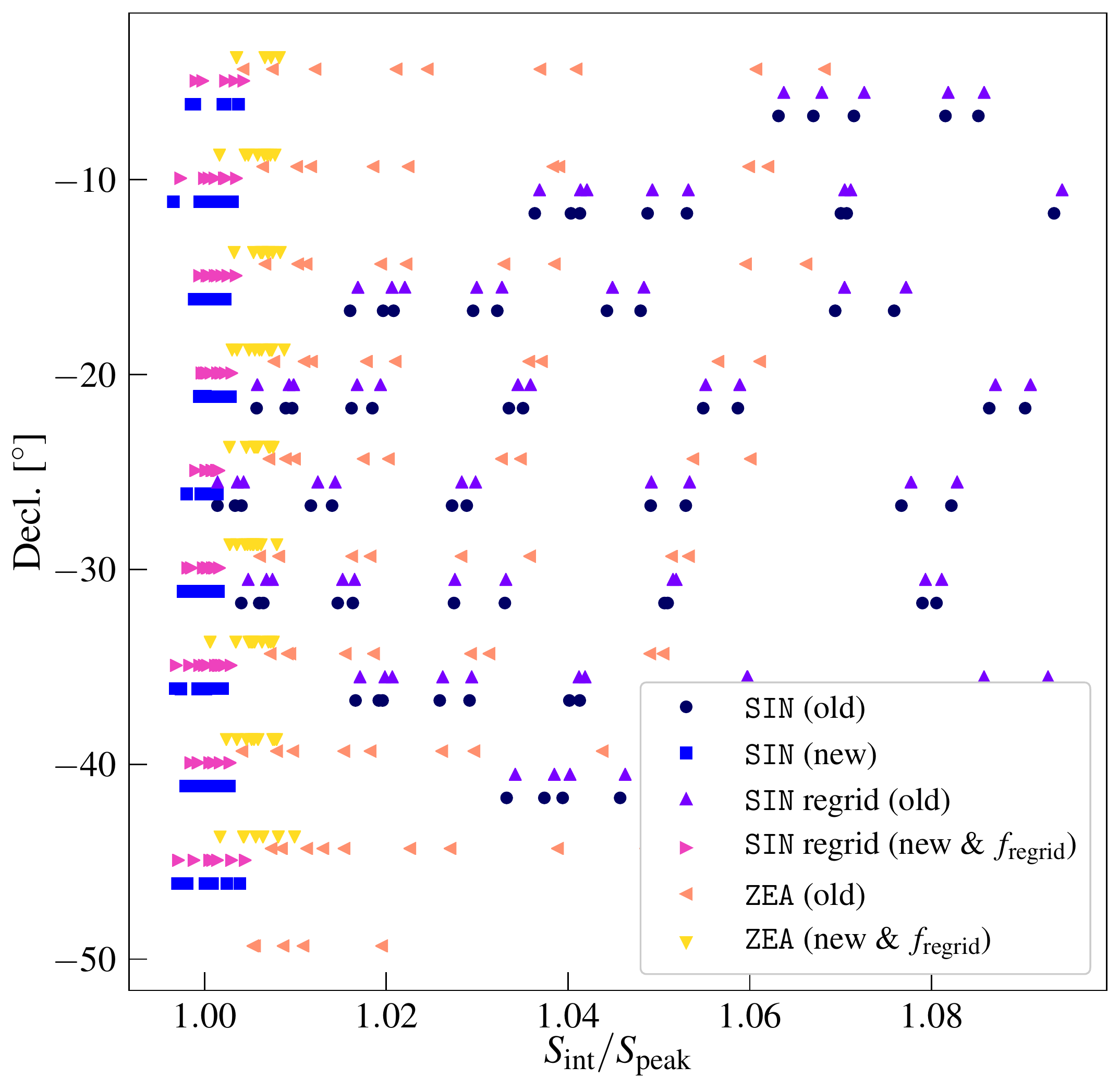}
    \caption{\label{fig:appendix:psf:zenith:dec} Zenith pointing, declination dependence.}
    \end{subfigure}\\%
    \caption{\label{fig:appendix:psf} The various effects on integrated flux density measurements. Note the datasets are shifted by an arbitrary 36\arcmin~for clarity. The \texttt{ZEA} projection is only used after reprojecting as the imaging software does not natively generate \texttt{ZEA} images. \subref{fig:appendix:psf:highdec:ra} and \subref{fig:appendix:psf:highdec:dec}: Low-elevation snapshot (from this work) where the snapshot is reprojected to a different set of coordinates, showing the RA and declination dependence, respectively. \subref{fig:appendix:psf:zenith:ra} and \subref{fig:appendix:psf:zenith:dec}: A zenith-pointed snapshot (used as a ``best-case'' example only) where the reprojection does not change the reference coordinate significantly, showing for the RA and declination dependence, respectively. Note the different scales of $S_\mathrm{int}/S_\mathrm{peak}$ between the two pointings.}
\end{figure*}

We demonstrate these two effects (general reprojection corrections and removal of PSF size calculations by \texttt{aegean}) by selecting two 88-MHz snapshots---a low-elevation snapshot (from this work) and a zenith-pointed snapshot (a ``best-case'' example)---to simulate grids of 1~Jy point sources across the field of view without noise. We then image the data using \texttt{wsclean} as with real data (including the use of a shift of phase centre to zenith), and reproject each resultant snapshot to an example set of coordinates that would be used when generating mosaics. Once images are prepared, we source-find with \texttt{aegean}: first with the old version of \texttt{aegean}, then without the internal PSF calculations and using $f_\mathrm{regrid}$ for the reprojected images. Fig. \ref{fig:appendix:psf} shows the results of the source-finding on the various images (original \texttt{SIN}, reprojected \texttt{SIN}, and reprojected \texttt{ZEA}). Note that a $<1$ per cent error remains after the correction, however, it is likely this falls within the expected error from the interpolation done during the reprojection process. Visual inspection of simulated point sources makes it clear that $f_\mathrm{regrid}$ applied to the PSF major axis mimics the reprojected point sources. \par

Post commit \texttt{6cd5bac}, \texttt{aegean} {does not} attempt to calculate the size of the PSF, leaving the user to supply an appropriate PSF map if needed \footnote{This is more appropriate than \texttt{aegean} trying to determine direction cosines with no knowledge of the original projection.}, and \texttt{python} code is available in \texttt{piip} (see Section \ref{sec:data:mwa}) to generate PSF maps with $f_\mathrm{regrid}$ applied. This is done as part of the mosaicking for this work. {Note that at the time of writing work is being done on \texttt{aegean} (from February 2020) to incorporate correct calculations of PSF and pixel sizes across an image for \texttt{SIN} projection images. From February 2020 up to commit \texttt{d453938} the $f_\mathrm{regrid}$ factors derived for \texttt{SIN} PSF maps are identical to those derived for \texttt{ZEA} and $f_\mathrm{regrid}$ for \texttt{ZEA} re-gridded images remains the same.}

\section{MWA flux scale corrections}\label{fluxwarp}
After ensuring PSF-related effects are removed, there are still other issues that arise in real MWA data reduction that result in final image flux scales not being consistent with the input amplitude calibration model. The effect is largely only problematic at low-elevations, which leads to the suspicion that the primary beam model used in correcting the individual snapshots is not accurately defined for these low-elevation pointings. The individual snapshots have differing pointings and so the final primary beam correction is slightly different between them. To correct this effect, we use an in-house developed \verb|python| code \verb|flux_warp| \footnote{\url{https://gitlab.com/Sunmish/flux_warp}} to \emph{finalise} the primary beam correction. The basic premise of \verb|flux_warp| is to take an image, image catalogue with measured flux densities, and a model catalogue of the sky to compare to, then create a screen to multiply the image by to correct, for example, primary-beam related problems. In principle a variety of model sky catalogues can be used but for this work we use the same model sky catalogue used for calibration, without bright extended sources (e.g. Virgo~A). The screen can be created using a number of methods: \begin{enumerate}
    \item (SNR-weighted) mean or median, \\
    \item (SNR-weighted) 1-D polynomial fit to declination or elevation, \\
    \item (SNR-weighted) 2-D polynomial fit to image pixel coordinates, or \\
    \item Interpolation using linear radial basis function, pure 2-D linear, or nearest-neighbour methods.
\end{enumerate}
While the beam effects appear elevation-dependent, we find that for these data this fitting does not reduce residuals as well as a linear radial basis function (RBF) interpolation (see \verb|scipy.interpolation.Rbf| \footnote{\url{https://docs.scipy.org/doc/scipy/reference/generated/scipy.interpolate.Rbf.html}}; \citealt{scipy}) method, thus we use this RBF method to determine appropriate flux-scale corrections to apply over the individual snapshots. For each snapshot, a number of ``calibrator'' sources are chosen satisfying \begin{equation}
    S_{\mathrm{cal},\nu} \geq 1~\mathrm{Jy} \left(\dfrac{\nu}{88~\mathrm{MHz}}\right)^{-0.77},
\end{equation}
where $\nu$ is effective frequency of the image and $S_{\mathrm{cal},\nu}$ is the flux density of the source. Additionally, we impose a constraint that only 1000 sources may be selected with the brightest sources preferentially chosen. This source number limit is largely due to computational time constraints. The exact number of ``calibrators'' chosen for each snapshots varies between 50 and 1000, with the higher frequency bands typically on the lower side. Of the 50--1000 calibrator sources initially selected, 25 per cent of these from the faint end of the set are reserved for testing the model and are not used in determining the model. 

\begin{figure}
    \centering
    \includegraphics[width=1\linewidth]{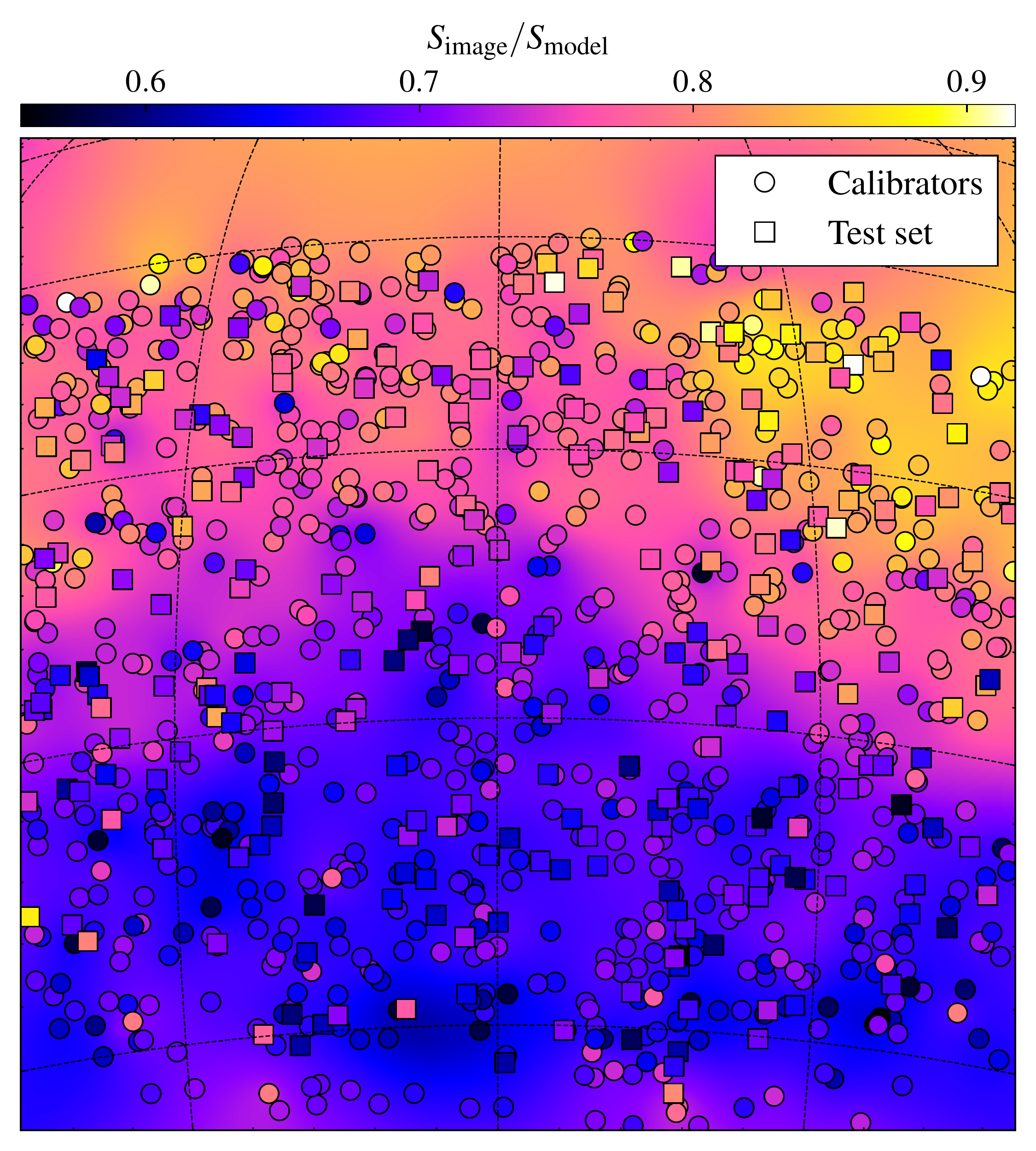}
    \caption{Example output showing the derived RBF interpolated correction factor map that is applied to the snapshot image with calibrator and test sources overlaid. The colour scale for the map and sources is the same. Note this image represents the full imaged region at 88~MHz which is $\sim44\degr \times 44\degr$.}
    \label{fig:fluxwarp:map}
\end{figure}

\begin{figure}
    \centering
    \includegraphics[width=1\linewidth]{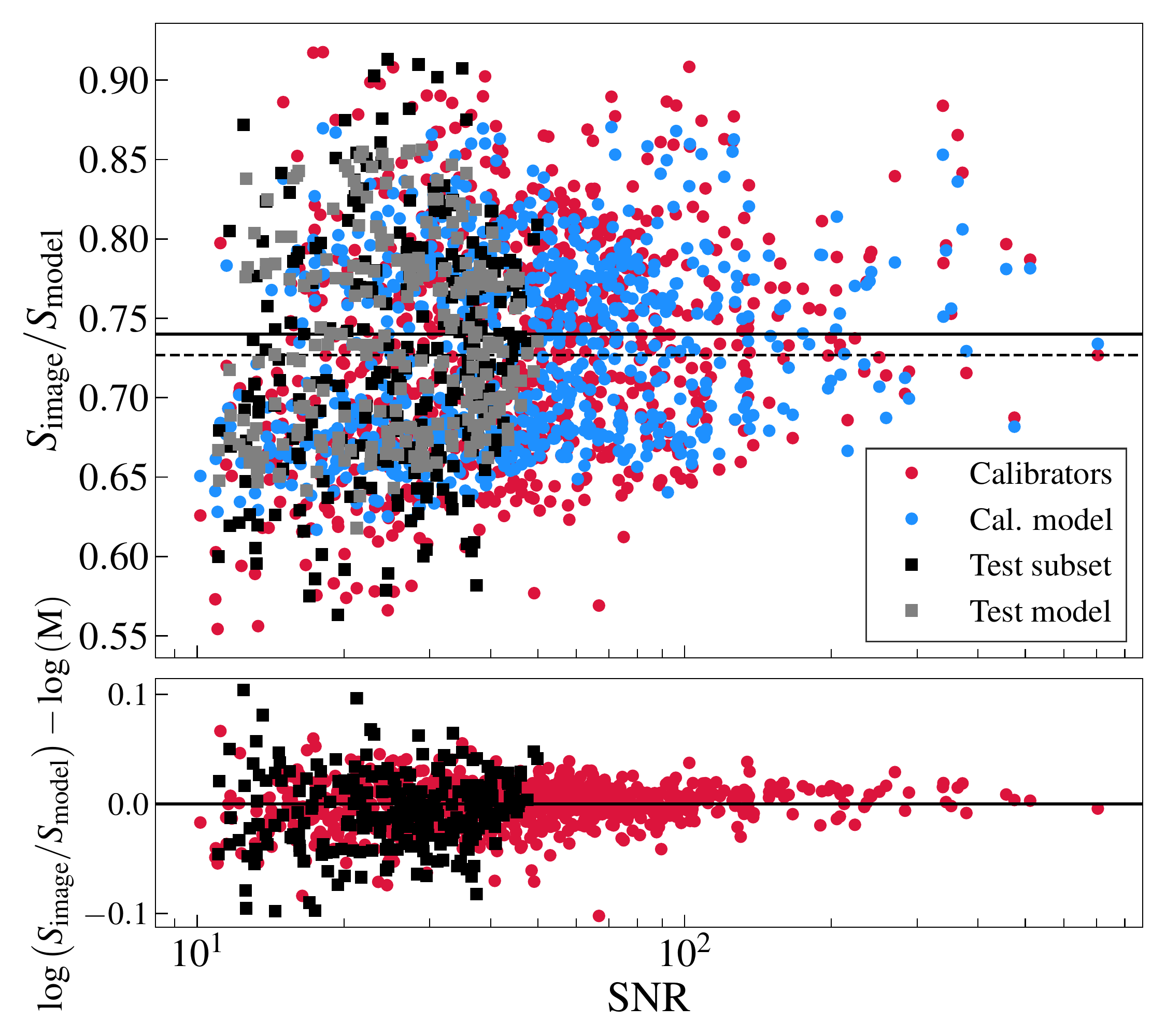}
    \caption{Example output showing the residuals between the calibrator and test sources when inspecting their new measured flux densities after applying the correction factor map, where M is the model factor, and the black solid and dashed lines are the mean and median factors, respectively.}
    \label{fig:fluxwarp:residuals}
\end{figure}

\begin{figure}
    \centering
    \includegraphics[width=1\linewidth]{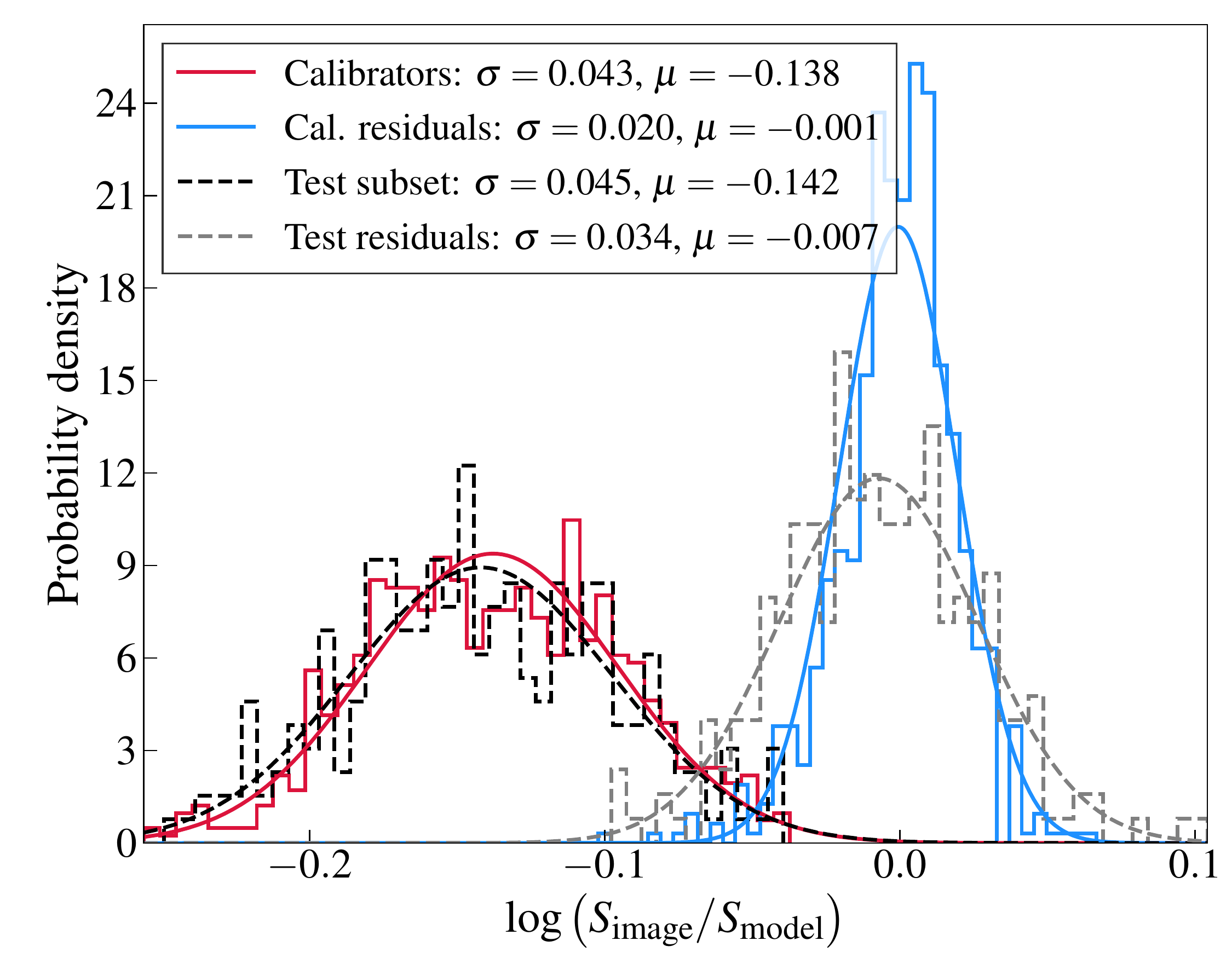}
    \caption{Example output showing the histogram of the log-ratios and their residuals after creating the correction factor map.}
    \label{fig:fluxwarp:histogram}
\end{figure}

Fig.~\ref{fig:fluxwarp:map} shows the derived correction factor map (where corrected data is the original data divided by the correction factor map) with the calibrator and test sources overlaid. Fig.~\ref{fig:fluxwarp:residuals} shows the residuals of the calibrator and test sources at their locations on the correction factor map. For this particular snapshot example (Obs.~ID 1200252120), the flux density threshold for calibrator sources was moved to 2.14~Jy (with maximum flux density 129~Jy) for 750 calibrators, and the flux density range for the 250 test sources was $1.77~\mathrm{Jy} < S < 2.14$~Jy. During a run of \verb|flux_warp|, a number of basic statistics are computed prior to creating the correction factor screen including fitting a normal distribution to the log-ratios (i.e. $\log\left[S_\mathrm{image} / S_\mathrm{model}\right]$). Fig.~\ref{fig:fluxwarp:histogram} shows this fitting to the log-ratios and residuals showing the improvement in the calibrators and test sources. \par
One final use of the \verb|flux_warp| is performing quality assurance on the stacked mosaics, where we determine the standard deviation of the measured integrated flux densities from our input model to estimate the intrinsic uncertainty in our absolute flux calibration. This results in attributing a few per cent flux scale error to each mosaic (see Table \ref{tab:phaseIIobs}).

\section{CLEAN bias}\label{appendix:bias}

{In comparing the \mwa~data to the TGSS data in the same region, we found that integrated flux densities were biased towards lower values. In the low-SNR case, this bias pushed the integrated flux density well below the peak flux in the map. We consider this (at least in part) due to CLEAN bias \citep[see e.g.][]{Becker95,White1997,Condon1998}, though note that the observed bias will have some contribution from the inherent bias in measuring integrated flux densities without, e.g., Gaussian fitting (as in the case for measuring the flux density of the source in Abell~1127).}

{We correct this by fitting the offset $S_\mathrm{peak} - S_\mathrm{int}$ for compact sources ($S_\mathrm{int} / S_\mathrm{peak} < 1.2$) with a linear function of the form $S_\mathrm{bias,compact} = A\times\mathrm{SNR} + B$ for each \mwa~image. For measuring source flux densities, we use a floodfill approach out to $2\sigma_\mathrm{rms}$ rather than Gaussian fitting to mimic the technique used in measuring the diffuse cluster source. Gaussian fitting would hide the issue, as the integrated flux density of a Gaussian source is measured from its fitted peak flux, and major/minor axes and does not directly measure the pixel values.}

{Table \ref{tab:bias} reports the bias-correcting parameters. Note that for an extended source, we assume that the bias scales with fractional peak, i.e. $S_\mathrm{bias} = S_\mathrm{bias,compact} \times \left(S_\mathrm{int} / S_\mathrm{peak} \right)$, becoming worse for low surface brightness sources. Note that due to the convention used the final integrated flux density is defined as $S_\mathrm{int,corrected} = S_\mathrm{int} + S_\mathrm{bias}$.}

\begin{table}
    \centering
    \caption{\label{tab:bias} {Fitted bias correction parameters for each \mwa~band for $S_\mathrm{bias,compact} = A\times\mathrm{SNR} + B$.}}
    \begin{tabular}{ccc}
         \hline
         Band & $A$ & $B$  \\
         \hline
          & (mJy) & (mJy) \\
         \hline
         ~72--103 & $-1.2$ & 62.2 \\
         103--134 & $-0.54$ & 25.1 \\
         139--170 & $-0.36$ & 14.1 \\
         170--200 & $-0.23$ & 11.0 \\
         200--231 & $-0.34$ & 13.3 \\
         \hline
    \end{tabular}
\end{table}

\end{appendix}

\end{document}